\documentstyle[amssymb,11pt]{amsart}
\newcommand{\Z}{{\Bbb Z}}
\newcommand{\N}{{\Bbb N}}

\newcommand{\C}{{\Bbb C}}
\newcommand{\Q}{{\Bbb Q}}
\newcommand{\Ref}[1]{{$($\ref{#1}$)$}}
\newcommand{\bean}{\begin{eqnarray}}
\newcommand{\eean}{\end{eqnarray}}
\newcommand{\be}{\begin{displaymath}}
\newcommand{\ee}{\end{displaymath}}
\newcommand{\bea}{\begin{eqnarray*}}
\newcommand{\eea}{\end{eqnarray*}}
\newcommand{\g}{{{\frak g}\,}}
\newcommand{\bor}{{{\frak b}}}
\newcommand{\n}{{{\frak n}}}
\newcommand{\h}{{{\frak h\,}}}
\newcommand{\Id}{{\mbox{Id}}}
\newcommand{\ad}{{\mbox{ad}}}

\newenvironment{proof}{\noindent{\it Proof\/}:}{$\;\Box$}
\newenvironment{definition}{\par\vspace{.5\baselineskip}
\noindent{\bf Definition\/}:}{\par\vspace{.5\baselineskip}}
\newtheorem%
{thm}{Theorem}[section]
\newtheorem%
{proposition}[thm]{Proposition}
\newtheorem%
{lemma}[thm]{Lemma}
\newtheorem%
{lemmadef}[thm]{Lemma-Definition}
\newtheorem%
{corollary}[thm]{Corollary}
\newtheorem%
{conjecture}[thm]{Conjecture}
\newcommand{\End}{{\operatorname{End}}}
\newcommand{\Hom}{{\operatorname{Hom}}}

\newcommand{\tr}{{\operatorname{tr}}}
\title[Integrable Schroedinger operators]{Three formulas for eigenfunctions
of integrable Schr\"odinger operators}
\thanks{The authors were supported in part
by NSF grants DMS-9400841 and DMS-9501290 respectively.
}
\author{Giovanni Felder and Alexander Varchenko}
\address{
Department of Mathematics,
Phillips Hall, University of North Carolina at Chapel Hill,
Chapel Hill, NC 27599-3250, USA%
}
\email
{felder@@math.unc.edu,
varchenko@@math.unc.edu.
{\it World Wide Web home page:}
http://www.math.unc.edu/Faculty/felder
}
\date{November 1995}
\begin{document}
\maketitle
\begin{abstract}
We give three formulas for meromorphic eigenfunctions (scattering
states) of Sutherland's
integrable $N$-body Schr\"odinger operators and their
generalizations.
The first is an explicit computation of the Etingof--Kirillov traces
of intertwining operators, the second an integral representation
of hypergeometric type, and the third of Bethe ansatz type.
 The last two formulas are degenerations of elliptic formulas
obtained previously in connection with the
Knizhnik--Zamolodchikov--Bernard
equation. The Bethe ansatz formulas in the elliptic case
are reviewed and discussed in more detail here: Eigenfunctions
are parametrized by a ``Hermite--Bethe'' variety, a generalization
of the
spectral variety of the Lam\'e operator.
We also give the $q$-deformed version of our
first formula. In the scalar $sl_N$ case, this gives common eigenfunctions
of the commuting Macdonald--Rujsenaars difference operators.
\end{abstract}

\section{Introduction}
Let $\g$ be a simple complex Lie algebra, with a non-degenerate ad-invariant
bilinear form $(\ ,\ )$, and a fixed Cartan subalgebra $\h$. Let
$\Delta\subset\h^*$ be the set of roots of $\g$. For each $\alpha\in\Delta$ let
$e_\alpha\in\g$ be a corresponding root vector, normalized so that
$(e_\alpha,e_{-\alpha})=1$.

Suppose that $U$ is a highest weight representation of $\g$ with finite
dimensional zero-weight space ( the space of
vectors in $U$ annihilated by $\h$) $U[0]$. We consider in this paper the
differential operator
\begin{equation}\label{ham} H=-\triangle+\sum_{\alpha\in\Delta}{\pi^2\over
\sin^2(\pi \alpha(\lambda))}e_\alpha
e_{-\alpha}
\end{equation}
acting on functions on $\h$ with values in $U[0]$. The Laplacian
$\triangle$
is the operator
$\sum_\nu{\partial^2\over\partial\lambda_\nu^2}$ in terms of
coordinates $\lambda_\nu=(b_\nu,\lambda)$ for any orthonormal
basis $b_1,\dots,b_r$ of $\h$.
As remarked in $\cite{EK1}$, if $\g=sl_N$, and
$U$ is the  symmetric power $S^{pN}\C^N$
of the defining representation $\C^N$,
then $U[0]$ is one-dimensional and
this differential operator reduces to Sutherland's integrable
$N$-body Schr\"odinger operator \cite{S} in one dimension
with coupling constant $p(p+1)$.
 In suitable variables, $H$ is in this case proportional to
\begin{equation}\label{scalar}
H_S=-\sum_{j=1}^N{\partial^2\over\partial x_j^2}+p(p+1)\sum_{j\neq l}
V(x_j-x_l)
\end{equation}
with $V(x)=1/\sin^2(x)$ (trigonometric model) or
$V(x)=1/\sinh^2(x)$ (hyperbolic model). We refer to this special
cases as the scalar cases.

We consider in this paper eigenfunctions (functions $\psi$ with
$H\psi=\epsilon\psi$ for some $\epsilon\in\C$) of the form
$e^{2\pi i(\xi,\lambda)}f(\lambda)$ where
$f$ is meromorphic and regular as $\alpha(\lambda)\to
i\infty$, $\forall\alpha\in\Delta$ (a more precise definition is
given below). In the language of $N$-body Schr\"odinger operators these are
{\em scattering states} for the hyperbolic models. It turns out that
for generic $\xi$ the space $E(\xi)$ of such eigenfunctions is finite
dimensional and isomorphic to $U[0]$.

We give three formulas for these eigenfunctions. The first one
(Theorem \ref{first})
is an explicit cumputation of
 an expression of Etingof and Kirillov in terms
of traces of certain intertwining operators.
The reason that these combinations of traces can be computed explicitly
is that they turn out to be the same as
for the corresponding Lie algebra without Serre relations
(Prop.\ \ref{free}). This observation reduces the computation
of traces
to a tractable combinatorial problem. We also give the $q$-deformed
version of this formula (Theorem \ref{last}).
The second formula, an integral representation (Theorem
\ref{integral}),
 follows from the fact that eigenfunctions can be
obtained as suitable limits of solutions of the
 Kni\-zhnik--Za\-mo\-lo\-dchi\-kov--Ber\-nard
equation of conformal field theory, for which we gave
explicit solutions in \cite{FV}. The third formula (Theorem
\ref{Bethe}) is of the Bethe ansatz (or Hermite) type: one has an explicit
expression of a function depending on parameters $T\in\C^n$. This
function is an eigenfunction if the parameter lie on an algebraic
variety which we call Hermite--Bethe variety. We have a regular map $p$
from the Hermite--Bethe variety to $\h$, sending $T$ to the
corresponding value of $\xi$. Conjecturally, $p$ is a covering map and a
basis of $E(\xi)$ for generic $\xi$ is obtained by taking the
eigenfunctions corresponding to the points in the fiber $p^{-1}(\xi)$.
This conjecture is proved in some cases including the scalar case
(Theorem \ref{cBS}).

The three formulas hold for generic values of the spectral parameter
$\xi$. In Section \ref{sec6} we study trigonometric polynomial
solutions of the eigenvalue problem $H\psi=\epsilon\psi$. These
(Weyl antiinvariant)
eigenfunctions are related to multivariable Jacobi polynomials
(or Jack polynomials). Formulas for these polynomials are obtained
as the spectral parameters tends to an antidominant integral weight.
The construction involves the construction of the
{\em scattering matrices} for the problem, which give a solution
of the Yang--Baxter equation.
It would be interesting to generalize this construction to the $q$-%
deformed case, which would give formulas for Macdonald polynomials.
We give such formulas in Section \ref{quantum} as a conjecture.
This conjecture can be proved by the same method as in the classical
case for  $sl_2$ and $sl_3$, but the regularity property of Weyl
antiinvariant eigenfunction appears to be more difficult to
prove in the general case.

The fact that we have three different formulas for the same
thing can be explained in informal terms as follows.
Our second formula is an integral depending on a complex
parameter $\kappa
\neq 0$.
Its form is
\be
\psi(\lambda)=e^{2\pi i\xi(\lambda)}\int_{\gamma(\kappa)}
\Phi_0(T)^{1/\kappa}\omega(T,\lambda)
\ee
The function $\Phi_0$ is a many-valued homolomorphic function on
$D_n=(\C-\{0,1\})^n-\cup_{i<j}\{T\,|\,T_i=T_j\}$ and
$\omega(T,\lambda)$ is a rational $U[0]$-valued differential
$n$-form on $D_n$. The cycle $\gamma(\kappa)$ has coefficients
in the local system determined by $\Phi_0(T)^{1/\kappa}$. It
turns out that, for all $\kappa$, $\gamma(\kappa)$,
 $\psi$ is an eigenfunction
of $H$ with the same eigenvalue $4\pi^2(\xi,\xi)$, and that (if $\xi$ is
generic) for any fixed generic $\kappa$
all eigenfunctions can be obtained by suitable $\gamma(\kappa)$.
 One can thus consider suitable families of cycles parametrized by $\kappa$
in the limits  $\kappa\to 0$, $\kappa\to\infty$. In the former
case the homology reduces to ordinary homology, and the answer is
given in terms of residues of $\omega$ (our first formula).
 In the latter limit, the integral
can be evaluated by the saddle point method and the eigenfunctions
are given by evaluating $\omega$ at the critical points of $\Phi_0$
(our third formula).

This reasoning turns out to be useful to write down the formulas,
(and to ``understand'' them)
but it is then easier to prove them by other methods.

In Section \ref{ellcase}, we generalize our results on
Bethe ansatz eigenfunctions (the third formula) to the elliptic case.
The Schr\"odinger operator in this case has the Weierstrass
$\wp$-function as a potential:
\begin{equation}\label{eham}
H_e=-\triangle+\sum_{\alpha\in\Delta}{\wp(\alpha(\lambda))}e_\alpha
e_{-\alpha}+\mbox{const}.
\end{equation}
In the scalar case $\g=sl_N$, $U=S^{pN}{\C^N}$, we obtain
\Ref{scalar} with $V=\wp$.
Bethe ansatz eigenfuctions of $H_e$ were given in \cite{FV}.
After reviewing this construction, we give a result on completeness
of Bethe states (Theorem \ref{eCBS})
parallel to the trigonometric case, in the scalar case and
the case of the adjoint representation of $sl_N$: for generic
values of the spectral parameter $\xi$ there exist $\dim(U[0])$
solutions of the Bethe ansatz equations corresponding to
linearly independent quasi-periodic eigenfunctions with
multiplier given by $\xi$.

The formulas discussed here in the trigonometric case
were obtained by considering the trigonometric
limit of solutions of the KZB equations for an elliptic
curve with one marked point. The same construction could be done
in the case of $N$ marked points. It turns out that in the
trigonometric limit one of the equations always reduces to an
eigenvalue equation for $H$. The new feature is that $U$ is
the tensor product of $N$ representations. The remaining commuting
operators define differential equations in the space of eigenfunctions
of $H$ with a fixed eigenvalue.

In the $q$-deformed case, we could only generalize the
first formula. It would be interesting to find the second and third formula
in the $q$-deformed case.

The organization of the paper is as follows: in Section \ref{sec2}
we define and describe the space of eigenfunctions we consider in
the trigonometric case. Three formulas for these eigenfunctions
are given in Section \ref{sec3}
 in terms of coordinates of singular
vectors, in Section \ref{sec4} as integrals, and in Section \ref{sec5}
by a Bethe ansatz.
In Section \ref{sec6}, the action of the Weyl group is discussed
along with the relation to multivariable Jacobi polynomials at
special values of the spectral parameter. In Section \ref{ellcase}
we discuss the Bethe ansatz in the elliptic case. Section
\ref{quantum} gives a generalization of the first formula to
the $q$-deformed case. The proof of the first formula is
contained in Section \ref{proofs}.

We conclude this introduction by fixing some notations and conventions.
The Cartan subalgebra $\h$ will be often identified with its dual
via $(\ ,\ )$.
We fix a set of simple roots $\alpha_1,\dots,\alpha_r\in\Delta$, and
write $Q$ for the root lattice $\oplus_i\Z\alpha_i$. Its positive part
$\oplus_i\N\alpha_i$, with $\N=\{0,1,\dots\}$, will be denoted by
$Q_+$. We denote by $\rho$ half the sum of the positive roots.
We will use the partial ordering $\beta\geq\beta'$ iff
$\beta-\beta'\in Q_+$ on $\h$ and write $\beta>\beta'$ if
$\beta\geq\beta'$ but $\beta\neq\beta'$. We set $\Delta_+=\Delta\cap Q_+$.
If $\beta\in Q$, we write $|\beta|^2=(\beta,\beta)$. We will
also use the notation $|A|$ to denote the cardinality of a set
$A$.  The group of permutations of $n$ letters is denoted by $S_n$.

\section{The $\psi$ functions}\label{sec2}
We introduce a space of meromorphic functions on $\h$ that is preserved
by the differential operator $H$.  For $\beta\in\h^*$, let
\begin{equation}\label{eq1}
X_\beta(\lambda)=e^{-2\pi i\beta(\lambda)},
\end{equation}
and write $X_\beta=X_i$ if $\beta=\alpha_i$.
Let $A$ be
the algebra of functions on $\h$ which can be represented
as meromorphic functions of $(X_1,\dots,
X_r)\in\C^r$ with poles belonging to set
$\cup_{\alpha\in\Delta}\{X_\alpha=1\}$.
For instance, if $\alpha$ is a positive root, the functions
\begin{equation}\label{geo}
{1\over \sin^2(\pi \alpha(\lambda))}=
-{4X_\alpha\over (1-X_\alpha)^2},
\qquad
\cot(\pi \alpha(\lambda))= i{1+X_\alpha\over 1-X_\alpha}
\end{equation}
belong to $A$. The algebra $A$ is a subalgebra of
$\hat A=\C[[X_1,\dots,X_r]]$

If $\xi\in\h^*$, we introduce the $A$-module $A(\xi)$ and the $\hat
A$-module $\hat A(\xi)$ of functions of the form $\exp(2\pi i\xi) f$
where $f\in A$, resp.\ $f\in \hat A$. These modules are preserved by
derivatives with respect to $\lambda$. Therefore $H$ preserves the
spaces $A(\xi)\otimes U[0]$, $\hat A(\xi)\otimes U[0]$.

Thus any $\psi\in\hat A(\xi)\otimes U[0]$ has the form
\be
\psi(\lambda)=\sum_{\beta\in Q}X_{\beta-\xi}(\lambda)\psi_\beta,
\ee
with $\psi_\beta\in U[0]$ vanishing if $\beta\not\in Q_+$.

\begin{thm}\label{p1}
For generic $\xi\in\h^*$, and any non-zero $u\in U[0]$, there exists
a unique $\psi=\sum_\beta X_{\beta-\xi}\psi_\beta
\in \hat A(\xi)\otimes U[0]$, such
that $H\psi=\epsilon\psi$, for some $\epsilon\in\C$, and such that
$\psi_0=u$.
Moreover, $\epsilon=(2\pi )^2(\xi,\xi)$, and $\psi\in A(\xi)\otimes U[0]$.
\end{thm}
\noindent
The existence and uniqueness of $\psi$ as a formal power
series was proved in the scalar case by Heckman and Opdam
(see Sect.\ 3 of \cite{HOI}), who generalized a classical
construction of Harish-Chandra (see \cite{He}, IV.5). The
fact that the series converges to a meromorphic function seems
to be new.

We next prove this theorem except for the statement
that the formal power series $\psi$ actually belongs to $A(\xi)\otimes U[0]$,
which will follow from the explicit formulas for $\psi$ given
below.

\begin{proof} The idea is that the eigenvalue equation
$H\psi=\epsilon\psi$ is a recursion relation for the coefficients
$\psi_\beta$.
We use the fact that $\alpha$ and $-\alpha$ give the same
contribution to the sum in $H$, to apply \Ref{geo}. With
the formula $\triangle X_{\beta-\xi}(\lambda)=
(2\pi i)^2(\beta-\xi,\beta-\xi)X_{\beta-\xi}(\lambda)$, we see that
$H\psi=\epsilon\psi$ is
equivalent to the recursion relation
\begin{equation}\label{eqrec}
[(\beta-\xi,\beta-\xi)-(2\pi)^{-2}\epsilon]\psi_\beta
=\sum_{j>0}2j \sum_{\alpha\in\Delta_+}
e_\alpha e_{-\alpha}\psi_{\beta-j\alpha}
\end{equation}
for the
coefficients $\psi_\beta\in U[0]$. The sum on the right-hand
side has only finitely many non-zero terms. The initial
condition for this recursion is $\psi_0=u\neq 0$. For
$\beta=0$, the equation reads then $(\xi,\xi)-(2\pi)^{-2}\epsilon=0$.
Thus there is a solution only if $\epsilon=(2\pi)^2(\xi,\xi)$.
For generic $\xi$, the coefficient of $\psi_\beta$ does
not vanish if $\beta\neq 0$, so \Ref{eqrec}
gives $\psi_\beta$ in terms of $\psi_{\beta'}$ with
$\beta'<\beta$. We conclude that, for generic $\xi$,
\Ref{eqrec} has a solution
if and only if $\epsilon=(2\pi)^2(\xi,\xi)$, and this solution is
unique.
\end{proof}

\begin{definition} We denote by $E(\xi)$ the vector space
of functions $\psi\in A(\xi)\otimes U[0]$ such that $H\psi=
(2\pi)^2(\xi,\xi)\psi$.
\end{definition}

\noindent{\bf Remarks.}
\noindent 1. It looks as if the definition of $E(\xi)$ should
depend on the choice of simple roots, but this is not so. From
the explicit expressions given below it follows that, for any
set $R$ of simple roots,
the functions in $E(\xi)$ can be analytically continued to
functions in $A_R(\xi)\otimes U[0]$, where $A_R(\xi)$ is the
space $A(\xi)$ defined using $R$. See Section \ref{sec6}.

\noindent 2. $H$ is part of a commutative $r$-dimensional
algebra $D$ of Weyl-invariant differential operators whose symbols
are (Weyl-invariant) polynomials on $\h^*$. These operators have
coefficients which are polynomials in $\cot(\pi\alpha(\lambda))$,
$\alpha\in\Delta_+$. It follows that $A(\xi)$ is preserved by $D$,
see \Ref{geo},
 and that the functions $\psi$ of the theorem are
 common eigenfunctions of all operators in $D$. Indeed if $L\in D$
with symbol $P$, then $L$ preserves $E(\xi)$ since it commutes with
$H$. If $\psi\in E(\xi)$ has leading term
$e^{2\pi i\xi(\lambda)}u$, $ u\in U[0]$, then $L\psi$ has leading term
 $P(2\pi i\xi)e^{2\pi i\xi(\lambda)}u$. Thus $L\psi=P(2\pi i\xi)\psi$.

\noindent 3. If $\g$ is a general Kac-Moody Lie algebra and
$U$ is a $\g$-module with finite dimensional zero-weight
space $U[0]$, $H$ is still a well-defined  endomorphism
of $\hat A(\xi)\otimes U[0]$: the (possibly infinite) sum over
$\Delta_+$ gives only finitely many contributions in each
fixed degree, as $\sin^{-2}(\pi\alpha(\lambda))= O(X_\alpha)$,
see \Ref{geo}. The above theorem holds
(except for the statement that $\psi\in A(\xi)\otimes U[0]$)
with the same proof.

\medskip
Etingof and Kirillov \cite{EK1} gave a representation theoretic construction
of eigenfunctions:
given $\xi$ generic and $u\in U[0]$, let $M_{\xi-\rho}$ be the
Verma module with highest weight $\xi-\rho$. Then there exists
a unique homomorphism $\Phi_u\in\Hom_\g(M_{\xi-\rho},M_{\xi-\rho}\otimes U)$,
such that the generating vector $v_{\xi-\rho}\in M_{\xi-\rho}$ is mapped to
$v_{\xi-\rho}\otimes u+\cdots$, up to terms whose first factor is
of lower weight. Then $\psi\in\hat A(\xi)\otimes U[0]$ is the ratio of
formal power series
\begin{equation}\label{eqtraces}
\psi_u(\lambda)=
{\tr_{M_{\xi-\rho}}\Phi_u \exp(2\pi i\lambda)
\over \tr_{M_{-\rho}}\exp(2\pi i\lambda)}.
\end{equation}
The traces are formal power series whose coefficients are
traces over the finite dimensional weight spaces of the Verma
modules. By definition, the trace of a map $M\to M\otimes U$
is the canonical map (for finite dimensional $M$)
\be
\tr_M:\Hom_{\C}(M,M\otimes U)\simeq
M^*\otimes M\otimes U\to U.
\ee
The numerator in \Ref{eqtraces} belongs to $\hat A(\xi-\rho)\otimes U[0]$,
and the denominator to $\hat A(-\rho)$ with leading coefficient
$1$, so the ratio is a well-defined element of $\hat A(\xi)\otimes U[0]$.
Combining this result with Theorem \ref{p1}, we get

\begin{proposition} For any generic $\xi\in\h^*$, the map
$u\to\psi_u$ defined by \Ref{eqtraces}
 is an isomorphism from $U[0]$ to $E(\xi)$.
\end{proposition}

\noindent{\bf Remark.}
\noindent Etingof and Styrkas \cite{ES} showed that
\Ref{eqtraces} viewed as a function of $\xi$, coincides
the Chalykh-Veselov $\psi$-function (see \cite{CV}, and \cite{ES}
for the matrix case considered in this paper), which is defined in terms of
its behavior as a function of $\xi$.

\section{The first formula}\label{sec3}
Our first formula is an explicit calculation of the trace of the
homomorphism $\Phi: M_{\xi-\rho}\to M_{\xi-\rho}\otimes U$ of the
previous section. The image of the generating vector is a singular
vector of weight $\xi-\rho$ (a vector of weight $\xi-\rho$
 killed by $e_\alpha$,
$\alpha\in\Delta_+$)
and all singular vectors of weight $\xi-\rho$ correspond to some
homomorphism.  Let $f_i=e_{-\alpha_i}$, $i=1,\dots,r$ and set
$f_I=f_{i_1}\cdots f_{i_m}$ for a multiindex $I=(i_1.\dots,i_m)$.  Let
$v_{\xi-\rho}\otimes u+\sum_If_Iv_{\xi-\rho}\otimes u_I$ be a singular
vector in $M_{\xi-\rho}\otimes U$ of weight $\xi-\rho$. Our formula
gives an eigenfunction in terms of the coefficients $u_I$.  Let us
remark that, for generic $\xi$, such a singular vector is uniquely
determined by its first coefficient $u\in U[0]$: the coefficients $u_I$
can be given rather explicitly in terms of $u$ and the inverse
Shapovalov matrix (see
\cite{ES}).

\begin{thm}\label{first}
 Let $\xi\in\h^*$  and let
 $v_{\xi-\rho}\otimes u+\sum_Lf_Lv_{\xi-\rho}\otimes u_L$
be a singular vector of weight $\xi-\rho$
in $M_{\xi-\rho}\otimes U$.  Then the function
$\psi(\lambda)=
e^{2\pi i\xi(\lambda)}(u+\sum_LA_L(\lambda)u_L)$, with
\begin{equation}\label{eqfirst}
A_{(l_1,\dots,l_p)}(\lambda)=\sum_{\sigma\in S_p}
\left(\prod_{j=1}^p
\frac
{X_{l_{\sigma(j)}}^{a_j+1}}
{1-X_{l_{\sigma(1)}}\cdots X_{l_{\sigma(j)}}}
\right)
f_{l_{\sigma(1)}}\cdots f_{l_{\sigma(p)}},
\end{equation}
where $a_j$ is the cardinality of the set of $m\in\{j,\dots,p-1\}$
such that $\sigma(m+1)<\sigma(m)$,
belongs to $E(\xi)$, and for generic $\xi$ all
functions in $E(\xi)$ are of this form.
\end{thm}
\noindent{\it Example.} Let $\g=sl_2$. If $U$ is irreducible, $U[0]$
is one-dimensional if $U$ has odd dimension and is zero otherwise.
Let $U$ be
a $2s+1$ dimensional
irreducible representation.
Our formula reduces to $\psi(\lambda)=e^{2\pi i\xi(\lambda)}
(u+\sum_{l=1}^s(\frac X{1-X})^lu_l)$, ($X=X_1$). The components $u_l$
of the singular vector are easily computed, and we get the formula
\begin{equation}\label{gamma}
\psi(\lambda)=e^{2\pi i\xi(\lambda)}
\sum_{l=0}^s
(-1)^l{(s+l)!\Gamma((\xi,\alpha)-l)
\over
l!(s-l)!\Gamma((\xi,\alpha))}
\left(\frac X{1-X}\right)^lu.
\end{equation}
We will prove the more general quantum version of Theorem \ref{first}
in section \ref{quantum}.

Note that Theorem \ref{first} completes the proof of Theorem \ref{p1}:
the coefficients are meromorphic functions. They seem
to have poles on $\{X_\beta=1\}$ for general $\beta\in Q$;
however, these poles cancel, since the differential equation
is regular there:

\begin{lemma}\label{regularity}
Let $\epsilon\in\C$ and suppose that $\psi$ is a meromorphic solution
of the differential equation
$H\psi=\epsilon\psi$ on $\h$, whose poles belong to the
union of the hyperplanes $H_{\beta,m}=\{\lambda|\beta(\lambda)=m\}$,
$\beta\in Q$, $m\in\Z$. Then $\psi$ is regular except possibly on
the hyperplanes $H_{\alpha,m}$, with $\alpha\in\Delta$, $m\in\Z$.
\end{lemma}
\begin{proof} Let $\beta\in Q-\Delta$, $m\in\Z$, and
 choose a system of affine coordinates $z_1,\dots,z_r$
on $\h$ so that $z_r=\beta(\lambda)-m$. If $\psi$ has
a pole of order $p>0$ on $H_{\beta,m}$, then, in the vicinity of
a generic point of $H_{\beta,m}$, $\psi=z_r^{-p}f(z_1,\dots,z_{r-1})
+\dots$ with non-vanishing regular $f$. The leading term of
$H\psi$ as $z_r\to 0$ comes from the Laplacian, as the potential
term is regular on $H_{\beta,m}$ and is equal
to $p(p+1)(\alpha,\alpha)z_r^{-p-2}f$ which is non-zero for $p>0$.
If $H\psi=\epsilon\psi$, then it follows that $\psi$ has a pole of order
$p+2$, a contradiction. We have shown that $\psi$ is regular
on generic points of $H_{\beta,m}$. Therefore, on any bounded
open subset $V$ of $\h$, the product of
$\psi$ by a suitable finite product of factors $\alpha(\lambda)-l$,
$\alpha\in\Delta$, $l\in\Z$ is holomorphic on the complement
of a set of codimension 2, and therefore everywhere on $V$ by
Hartogs' theorem.
\end{proof}

\noindent
This concludes the proof of Theorem \ref{p1}.

\section{The second formula}\label{sec4}
Our second formula is an integral representation. It is
obtained as the trigonometric limit of the integral representation
of solutions of the Kni\-zhnik--Za\-mo\-lo\-dchi\-kov--Ber\-nard (KZB)
 equations on elliptic curves with
one marked point.

One ingredient in the integrand is the rational function
of $2n$ variables $T_1,\dots,T_n$, $Y_1,\dots,Y_n$
\be
W(T,y)=
\prod_{j=1}^n\left({1\over T_j-T_{j+1}}-
{Y_1\cdots Y_j\over Y_1\cdots Y_j-1}\,{1\over T_j}\right),\qquad T_{n+1}:=1.
\ee
Now suppose that $U$ is an irreducible highest weight representation
with non-trivial zero-weight space. The highest weight
 of $U$ is then in $Q_+$, i.e., of the form
$\Lambda=\sum_jn_j\alpha_j$, for some non-negative integers $n_j$.
Set $n=\sum_jn_j$.
It is convenient to introduce the associated ``color''
 function $c:\{1,\dots,n\}
\to\{1,\dots,r\}$, the unique non-decreasing function with
$|c^{-1}(\{j\})|=n_j$, $j=1,\dots,n$.
For each permutation $\sigma\in S_n$, let $W_{\sigma,c}$
be the rational function of $n+r$ variables $T_1,\dots,T_n,
X_1,\dots,X_r$
\begin{equation}\label{W66}
W_{\sigma,c}(T,X)=W(T_{\sigma(1)},\dots, T_{\sigma(n)},
X_{c(\sigma(1))},
\dots,X_{c(\sigma(n))}).
\end{equation}
The other ingredient is the many-valued function of $T_1,\dots,T_n$,
\be
\Phi^\kappa_{\xi,\Lambda}(T)=\prod_{j<l}(T_j-T_l)^{(\alpha_{c(j)},\alpha_{c(l)})/\kappa}
\prod_{j=1}^n
T_j^{-{(\xi-\rho,\alpha_{c(j)})/\kappa}}(T_j-1)^{-{(\Lambda,\alpha_{c(j)})/\kappa}}.
\ee
We will consider integrals of
$\Phi^\kappa_{\xi,\Lambda}(T)W_{\sigma,c}(T,X)dT_1\cdots dT_n$ over
connected components   $\gamma$ of $\{t\in (0,1)^n|
t_i\neq t_j, (i\neq j)\}$. These ``hypergeometric integrals''
are defined as (meromorphic)
analytic continuation in the exponents of $T_i-T_j$,
$T_j$, $T_j-1$ in $\Phi^\kappa_{\xi,\Lambda}$ from a region in
which the integral converges absolutely. We say that a hypergeometric
integral {\em exists} if this analytic continuation is finite at
the given value of the exponents.
\begin{thm}\label{integral}
Suppose that $U$ is an irreducible
 highest weight module with highest weight $\Lambda
=\sum_jn_j\alpha_j$ and highest weight vector $v_\Lambda$.
 Set $n=\sum n_j$
and let $c:\{1,\dots,n\}\to\{1,\dots,r\}$
be the unique non-decreasing function such that $c^{-1}\{j\}$ has
$n_j$ elements, for all $j=1,\dots, r$.
Fix a generic complex number $\kappa$ and $\xi\in\h$ also generic.
Then, for each connected component $\gamma$ of $\{t\in (0,1)^n|
t_i\neq t_j, (i\neq j)\}$, the integral
\begin{equation}\label{eq00}
\psi_\gamma(\lambda)=
e^{2\pi i\xi(\lambda)}\int_\gamma
\Phi^\kappa_{\xi,\Lambda}(T)
\sum_{\sigma\in S_n}
W_{\sigma,c}(T,X(\lambda))\;dT_1\cdots dT_n\;
f_{c(\sigma(1))}\cdots f_{c(\sigma(n))}v_\Lambda,
\end{equation}
exists and defines a function in $E(\xi)$.
 Moreover, for each generic $\kappa$ and $\xi$,
all functions in $E(\xi)$ can be represented in this way.
\end{thm}

\noindent
In the rest of this section we prove this theorem.
The existence of the integral follows from \cite{V},
Theorem 10.7.12. Indeed,
the coefficients in \Ref{eq00} are linear combinations
of integrals considered in \cite{V} in the context of the
Knizhnik--Zamolodchikov equation with two points.

Clearly
$\psi_\gamma$  is of the form $e^{2\pi i\xi(\lambda)}$ times
a meromorphic function of $X_1,\dots,X_r$ with poles on
the divisors $X_\beta=1$, $\beta\in Q$. To prove that
$\psi_\gamma$ belongs to $E(\xi)$ it is therefore sufficient,
thanks to Lemma \ref{regularity}, to show that $\psi_\gamma$ is
an eigenfunction of $H$.

This follows from the results of \cite{FV}, which we now recall.

Let us define, for $q=\exp(2\pi i\tau)$ in the open unit
disk, $\theta(x)=\pi^{-1}\sin(\pi x)
\Pi_1^\infty(1-2q^j\cos(\pi x)+q^{2j})$ (in the notation
of \cite{FV}, $\theta(x)=\theta_1(x)/\theta'_1(0)$)
and $v(x)=-\frac {d^2}{dx^2}
\ln \theta(x)$. The function $v$ is doubly periodic with periods
1 and $\tau$.
 Then for any $\kappa\in\C-\{0\}$,
the  KZB equation is a partial differential equation for a
function $u(\lambda,\tau)$ on $\h\times\{\tau\in\C\,|\,\mbox{Im}\,\tau>0\}$,
with values in $U[0]$:
\begin{equation}\label{KZB}
4\pi i\kappa{\partial u\over\partial\tau}
=\triangle u-\sum_{\alpha\in\Delta}v(\alpha(\lambda))e_\alpha e_{-\alpha}u.
\end{equation}
As $q\to 0$, $v(x)\to\pi^2\sin^{-2}(\pi x)$, and the differential
operator on the right-hand
side converges to $-H$.
\begin{proposition}
Suppose that $u$ is a meromorphic solution of the KZB equation \Ref{KZB}
such that $q^{-a/2\kappa}u$ is a meromorphic  function of
$\lambda$ and $q$, $|q|<1$, and, as $q\to 0$,
\be
u(\lambda,\tau)=q^{a/2\kappa}(\psi(\lambda)+O(q))
\ee
Then $\psi$ obeys $H\psi=\epsilon\psi$, with $\epsilon=(2\pi)^2a$.
\end{proposition}
\noindent
The proof consists of comparing the leading coefficients in
the expansion of both sides of the KZB equation in powers
of $q$ at $q=0$.

A source of solutions of \Ref{KZB}  with the property described in the
Proposition is \cite{FV}. In that paper, we gave integral formulas
for solutions.

A class of solutions with asymptotic behavior as above is constructed
as follows (see \cite{FV} for more details).
Let $U$ be an irreducible highest weight module with highest
weight $\Lambda\in Q_+$, and let
as above $c:\{1,\dots,n\}\to\{1,\dots,r\}$ the non-decreasing
function associated to $\Lambda$.

Solutions are labeled by an element $\xi$ of $\h$ and a
connected component $\gamma$ of $\{t\in (0,1)^n|
t_i\neq t_j, (i\neq j)\}$. Let $\phi_{\xi,\Lambda}^\kappa$ be
a choice of branch over $\gamma$ of the many-valued function
\begin{equation}\label{ls}
e^{\pi i(\xi,\xi)\tau/\kappa+
2\pi i(\xi,\lambda+\kappa^{-1}\sum_jt_j\alpha_{c(j)})}
\prod_{i<j}\theta(t_i-t_j)^%
{(\alpha_{c(i)},\alpha_{c(j)})/\kappa}
\prod_{j=1}^n\theta(t_j)^{-(\alpha_{c(j)},\Lambda)/\kappa},
\end{equation}
and let $w$ be the meromorphic function of
$t_1,\dots,t_n$, $y_1,\dots,y_n$
\be
w(t,y)=
\prod_{j=1}^n{\theta(y_1+\cdots+y_j-t_j+t_{j+1})
\over\theta(y_1+\cdots+y_j)\theta(t_j-t_{j+1})}
\ee
and for each
permutation $\sigma\in S_n$, let $w_{\sigma,c}$
be the meromorphic function of $n+r$ variables $t_1,\dots,t_n,
x_1,\dots,x_r$
\begin{equation}\label{e67}
w_{\sigma,c}(t,x)=w(t_{\sigma(1)},\dots, t_{\sigma(n)},
x_{c(\sigma(1))},
\dots,x_{c(\sigma(n))}).
\end{equation}
Consider the
differential $n$-form  with values
in $U[0]$ depending on $\lambda\in\h$,
\be
\omega(\lambda,\tau)=\sum_{\sigma\in S_n}
w_{\sigma,c}(t,\alpha_1(\lambda),\dots,\alpha_r(\lambda))\;dt_1\cdots dt_n\;
f_{c(\sigma(1))}\cdots f_{c(\sigma(n))}v_\Lambda.
\ee
Then the integral
\be
\int_\gamma\phi_{\xi,\Lambda}^\kappa(t)\omega(\lambda,\tau)
\ee
exists (as analytic continuation in the exponents from a region
of absolute convergence)
and is a solution of the KZB equation.
As $\tau\to i\infty$,
\be
\theta(x)=\sin(\pi x)/\pi +O(q),
\ee
and the function \Ref{ls} behaves as
\be
q^{(\xi,\xi)/2\kappa}e^{2\pi i\xi(\lambda)}
[\Phi^\kappa_{\xi,\Lambda}(T)+O(q)]
\ee
in terms of the exponential variables $T_j=\exp(-2\pi it_j)$.
 Let $X_j=\exp(-2\pi i\alpha_j(\lambda))$.  The components of the
differential form $\omega$ behave as \be w_{\sigma,c}(t,x)dt_1\dots
dt_n=W_{\sigma,c}(T,X)dT_1\dots dT_n+O(q).  \ee Therefore, the
solutions corresponding to cycles in $H_n(C^0_n(\tau),\cal L(\xi))$
have the properties of the Proposition, with $a=(\xi,\xi)$, and $\psi$
is as stated in the claim of Theorem \ref{integral}.  We obtain in the
limit an integral representation for eigenfunctions as in the Theorem
but with an integration domain $\tilde\gamma_\sigma$ of the form
$0<\mbox{arg}(T_{\sigma(1)})<\cdots<\mbox{arg}(T_{\sigma(n)})<2\pi$
for some $\sigma\in S_n$. This integration domain may be deformed, so
that the corresponding integral can be written as linear combination
of integrals over the domains $\gamma_{\sigma'}$ defined by
$1>T_{\sigma'(1)}>\cdots>T_{\sigma'(n)}>0$. If we choose
$\xi=ia\xi_0$, for some $\xi_0$ such that $(\xi_0,\alpha_j)>0$ for all
$j$, and let $a$ tend to infinity, then the integral over
$\tilde\gamma_\sigma$ is equal to the integral over $\gamma_\sigma$
(for suitable choice of branch and orientation) plus terms that tend
to zero. It follows that, for generic $\xi$ and all $\sigma\in S_n$,
the integral over $\gamma_\sigma$ can be expressed as a linear
combination of integrals over $\tilde\gamma_{\sigma'}$ and is
therefore also an eigenfunction.

This completes the proof of the first part of Theorem \ref{integral}.

We must still prove that, in the generic case, all eigenfucntions
in $E(\xi)$ admit such an integral representation. In view of
Theorem \ref{p1}, it is sufficient to show that for every
$u\in U[0]$, there exists a cycle $\gamma$ such that
$e^{-2\pi i\xi(\lambda)}\psi_\gamma\to u$ as $X_j\to 0$
$j=1,\dots,r$. In other words, one must show that all vectors
in $U[0]$ are of the form
\be
\int_\gamma\Phi_{\xi,\Lambda}^\kappa(T)
\sum_{\sigma\in S_n}\bigl(\prod_{j=1}^{n-1}
\frac 1{T_{\sigma(j)}-T_{\sigma(j+1)}}\bigr)
\frac1{T_{\sigma(n)}-1}
f_{c(\sigma(1))}\cdots f_{c(\sigma(n))}v_\Lambda dT_1\cdots dT_n.
\ee
for some cycle $\gamma$. But this follows from the results in
\cite{V}, Theorem 12.5.5.

\section{The third formula}\label{sec5}

This is a formula of the Bethe ansatz type.
\begin{thm}
\label{Bethe}
Suppose $U$ is an irreducible
 highest weight module with highest weight $\Lambda
=\sum_jn_j\alpha_j$ and highest weight vector $v_\Lambda$.
 Set $n=\sum n_j$
and let $c:\{1,\dots,n\}\to\{1,\dots,r\}$
be the unique non-decreasing function such that $c^{-1}\{j\}$ has
$n_j$ elements, for all $j=1$,\dots, $r$.
Then the function parametrized by $T\in\C^n$
\begin{equation}\label{B66}
\psi(T,\lambda)=
e^{2\pi i\xi(\lambda)}
\sum_{\sigma\in S_n}
W_{\sigma,c}(T,X(\lambda))
f_{c(\sigma(1))}\cdots f_{c(\sigma(n))}v_\Lambda
\end{equation}
(see \Ref{W66} for the definition of $W_{\sigma,c}$)
belongs to $E(\xi)$ if the parameters $T_1$, \dots, $T_n$
are a solution of the set of $n$ algebraic equations
(``Bethe ansatz equations'')
\be
\bigl(\sum_{l:l\neq j}
{\alpha_{c(l)}\over T_j-T_l}-{\Lambda \over T_j-1}-
{\xi-\rho\over T_j},\alpha_{c(j)}\bigr)=0, \qquad j=1,\dots, n.
\ee
\end{thm}
\noindent{\bf Remark.} These Bethe ansatz equations are the same as
the Bethe ansatz equations of a special case of the Gaudin model
(cf.\ \cite{RV}).
The solutions are the critical points of the function
\be
\prod_{1\leq i\leq j\leq n}(T_i-T_j)^{(\alpha_{c(i)},\alpha_{c(j)})}
\prod_{j=1}^n
T_j^{-(\xi-\rho,\alpha_{c(j)})}
(T_j-1)^{-(\Lambda,\alpha_{c(j)})}.
\ee
\medskip
\noindent
Theorem \ref{Bethe}
 can be understood intuitively as a consequence of
Theorem \ref{integral}: one calculates the integrals, which up
to normalization are independent of $\kappa$, in the limit
$\kappa\to 0$, using the saddle point method.

The proof of this theorem can be taken directly from \cite{FV}
in the case of a degenerate elliptic curve.

The above result motivates the notion of Hermite--Bethe
variety. Let $\Lambda=\sum_jn_j\alpha_j$. Without loss
of generality, we may take $n_j>0$, for all $j=1,\dots,r$
(if this condition is not fulfilled, we may pass to a
subalgebra).
If $c:\{1,\dots,n\}\to\{1,\dots,r\}$ is the associated
non-decreasing function, we let $S_c$ be the product
of symmetric groups $S_{n_1}\times\cdots\times S_{n_r}$.
It acts on $\C^n$ by permutation of the variables $T_j$ with
same $c(j)$ and is a symmetry group of the system of Bethe
ansatz equations.
Moreover, we have, for all $\sigma\in S_c$,
\be
\psi(\sigma T,\lambda)=\psi(T,\lambda).
\ee
Let us write the Bethe ansatz equations as
$B_j(T)=0$, with
\be
B_j(T)=\bigl(\sum_{l:l\neq j}
{\alpha_{c(l)}T_j\over T_j-T_l}-{\Lambda T_j\over T_j-1}-
{\xi-\rho},\alpha_{c(j)}\bigr).
\ee
Subtracting pairs of equations we may eliminate $\xi$:

\begin{definition}
Let $F_n=(\C-\{0,1\})^n-\cup_{i<j:(\alpha_{c(i)},\alpha_{c(j)})\neq 0}
\{T|T_i=T_j\}$.
The Hermite--Bethe variety $H\!B(c)$ is
\be
H\!B(c)=\{T\in F_n| B_j(T)=B_{j+1}(T),
j\in\{1,\dots,n\}-\{n_1,n_1+n_2,\dots,n\}\}/S_c.
\ee
\end{definition}

The remaining equations define a regular map
$p:H\!B(c)\to \h$, $T\mapsto \xi$. The completeness hypothesis of
Bethe states is in this case:

\begin{conjecture}\label{conj} (cf.\ \cite{V1}, \cite{OT} and
conjectures in \cite{RV})
The map $p$ has dense image, and the generic fiber
consists of (at least) $\dim(U[0])$ points. The eigenfunctions
corresponding to points in $p^{-1}(\xi)$ span $E(\xi)$.
\end{conjecture}

In the case of $sl_2$ the conjecture holds and goes back
to Hermite, see \cite{WW}.

We prove this conjecture in two special situations, including the
scalar case relevant for many-body systems.

\begin{thm}\label{cBS}
Conjecture \ref{conj} holds in the following cases:
(a) $\g=sl_N$, $U=S^{pN}\C^N$, $p=1,2,\dots$, (see \Ref{scalar})
(b) $\g=sl_N$, $U=$ adjoint representation.
\end{thm}

\begin{proof} The proof is by induction in $N$, viewing
$sl_N$ as a Lie subalgebra of $sl_{N+1}$. We choose simple roots
$\alpha_1,\dots,\alpha_{N}$ of $sl_N$ in such a way that
$\alpha_1,\dots,\alpha_{N-1}$ are simple roots of $sl_{N-1}$. Also,
it is convenient to replace $\xi$ by $\zeta=\xi-\rho$.

 In case (a),
the highest weight of $U_N=S^{pN}\C^N$ is
$\Lambda=p\sum_{j=1}^{N-1}j\alpha_j$,  and $U_N[0]$ is one-dimensional.
The fiber over $\xi=\zeta+\rho$ of the Hermite-Bethe
 variety consists of the critical
points of
\be
\Phi_N(\zeta,T)=
\prod_{j<l}(T_j-T_l)^{(\alpha_{c(j)},\alpha_{c(l)})}
\prod_jT_j^{-{(\zeta,\alpha_{c(j)})}}
(1-T_j)^{-{(\Lambda,\alpha_{c(j)})}},
\ee
viewed as a function of $T=(T_1,\dots,T_{pN(N+1)/2})$. We have
$c(j)=m$ if $pm(m-1)/2<c(j)\leq pm(m+1)/2$, $1\leq m\leq N$.
We prove inductively that, for generic $\zeta$, $\Phi_N$ has a non-degenerate
critical point, and that the corresponding eigenfunction spans
$E(\zeta+\rho)$. Let us consider critical points of $\Phi_{N+1}(\zeta,T)$ when
$(\alpha_N,\zeta)=\epsilon^{-1}$ tends to infinity, and
$(\alpha_j,\zeta)$, $j<N$ are kept fixed. Let $\bar\xi$ denote the orthogonal
projection of $\xi$ onto the Cartan subalgebra of $sl_N$.  Let us
replace the coordinates $T_j$ indexed by $j$ such that $c(j)=N$ by new
coordinates $a_j$ defined by $T_j=1-\epsilon a_j$ ($c(j)=N$), and let
$\bar T=(T_1,\dots,T_{pN(N+1)/2})$ denote the remaining coordinates.
Then
\begin{eqnarray}\label{e66}
\epsilon^{p(p+1)N}\Phi_{N+1}(\xi,T)
&\!\!=\!\!&\Phi_N(\bar\zeta,\bar T)\prod_{c(j)=N-1}(1-T_j)^{pN}
\prod_{c(j)=N-1,c(l)=N}(T_j-1+\epsilon a_l)^{-1}\notag
\\ &\times\!&
\prod_{c(l)=N}(1-\epsilon a_l)^{-1/\epsilon}
a_l^{-{p(N+1)}}\prod_{j<l}(a_l-a_j)^2.
\end{eqnarray}
This function converges as $\epsilon\to 0$ to
a constant times
\be
\Phi_N(\bar\zeta,\bar T)
\prod_{l}e^{ a_l}
a_l^{-{p(N+1)}}\prod_{j<l}(a_l-a_j)^2.
\ee
The factor depending on the $a_j$'s has a non-degenerate critical point
$a_*$
in the domain $a_j\neq a_k\neq 0,$ $(j\neq k)$. This follows from the
fact that this factor is obtained as a limit $M\to\infty$ of
\be
\prod_{j}(1-a_j/M)^{-M}a_j^{-p(N+1)}\prod_{j<l}(a_j-a_l)^2.
\ee
The critical points of the latter function are known explicitly
(see (1.3.1) and (1.4.2) in \cite{V1}),
and it can be easily checked that they have
a limit as $M\to\infty$ in the domain $a_j\neq a_k\neq 0$, which
is therefore a critical point of the limiting function.
In fact, $a_*$ is the set of zeros of the polynomial
solutions of the differential equation $xy''+(x-m)y'-Ny=0$,
where $m=p(N+1)$.
By the induction hypothesis, $\Phi_N(\bar \zeta,\bar T)$ has a
non-degenerate critical point $\bar T_*$. Moreover, in a neighborhood
of $(a_*,\bar T_*,\epsilon=0)$, the right-hand side of \Ref{e66}
is holomorphic, which implies that the non-degenerate critical point
at $\epsilon=0$ deforms to a non-degenerate critical point for
generic $\epsilon$.

It remains to show that the corresponding eigenfunction generically spans
the one-dimensional vector space $E(\zeta+\rho)$, i.e., that it does not
vanish. Assuming this inductively to hold for $sl_N$, we see that
as $\epsilon\to 0$, the leading contribution in the sum \Ref{B66} is
given by permutations such that $c(\sigma(j))=N$, whenever $c(j)=N$.
These permutations give terms proportional to $f_{c(\tau(1))}\cdots
f_{c(\tau(pN(N+1)/2))}(f_N)^{pN}v_\Lambda$, for some $\tau\in S_{pN(N+1)/2}$.
It follows that when $\epsilon\to 0$ and $T(\epsilon)$ is
the above family of critical points,
$e^{-2\pi i(\xi,\lambda)}\psi_{sl_{N+1}}(T(\epsilon),\lambda)$
 converges
to $C(a_*)e^{-2\pi i(\bar\xi,\bar\lambda)}
\psi_{sl_N}(\bar T_*,\bar\lambda)$,
where the representation $U_N=S^{pN}\C^N$ of $sl_N$ is viewed as the
$sl_N$-submodule of $U_{N+1}$ generated by the singular vector
$(f_N)^{Np}v_\Lambda$. Using the identity
\be
\sum_{\sigma\in S_N}\frac{1}{(a_{\sigma(1)}-a_{\sigma(2)})\cdots
(a_{\sigma(N-1)}-a_{\sigma(N)})a_{\sigma(N)}}=
\frac1{a_1\cdots a_N},
\ee
we see that $C(a_*)$ is, up to a trivial nonzero factor, the inverse
of  the product
of the components $a_j$ of $a_*$, and therefore nonzero. This completes
the proof of part (a) of the theorem.

 The case (b) is treated in a similar way. The highest weight
of the adjoint represnetation $U_N$ of $sl_N$
is $\Lambda=\sum\alpha_i$ and  $c=\Id$.
The function $\Phi$ for $sl_N$ is
\be
\Phi_N(\zeta,T)=
\prod_{j=1}^{N-2}(T_j-T_{j+1})^{-1}
\prod_{j=1}^{N-1}T_j^{-{(\zeta,\alpha_{j})}}
(1-T_1)^{-1}(1-T_{N-1})^{-1}
\ee
As before we let
$\epsilon^{-1}=(\zeta,\alpha_N)$
go to infinity. If we set $T_{N}=1-\epsilon a$, $\bar
T=(T_1,\dots,T_{N-1})$, $\bar \zeta=$ projection of $\zeta$ onto the
Cartan subalgebra of $sl_{N}$,
then
\be
\epsilon\Phi_{N+1}(\zeta,T)
=\Phi_{N}(\bar\zeta,\bar T)\frac{1-T_{N-1}}{T_{N-1}-1+\epsilon a}
(1-\epsilon a)^{-1/\epsilon}
a^{-1}\buildrel{\epsilon\to 0}\over{\longrightarrow}
-\Phi_N(\bar\zeta,\bar T)\frac{e^a}a.
\ee
Proceeding as before, we see that non-degenerate critical points $\bar
T_*$, $a_*=1$ of the limiting function deform to non-degenerate
critical points for any generic $\epsilon$. the corresponding
eigenfunctions
deform to eigenfunctions of $sl_{N}$ taking values in $U_{N}[0]$,
viewed as a subspace of $U_{N+1}[0]$ via the inclusion of $sl_{N}$ in
$sl_{N+1}$. In this way we get $N-1$ linearly independent eigenfunctions
in the $N$-dimensional space $E(\zeta+\rho)$ associated to the adjoint
representation
of $sl_{N+1}$. To find the remaining eigenfunction, we let
$T_j=1-\epsilon a_j$ for all $j$. Then as $\epsilon\to 0$,
$\epsilon^m\Phi_{N+1}$ for suitable $m$ converges to
\be
\frac{e^{a_N}}
{a_1a_{N-1}}
\prod_{j=1}^{N-1}(a_{j+1}-a_j)^{-1}
\ee
The critical point of this function can be computed explicitly:
$a_j=j(1+1/N)$. It deforms to a non-degenerate critical
point for $\epsilon$ generic. The corresponding eigenfunction converges,
as $\epsilon\to 0$, to a constant function. From its explicit expression
it is clear that its value is not in $U_{N}=U(sl_N)f_Nv_\Lambda$
 and is therefore linearly
independent from the $N-1$ constructed before.
\end{proof}

\noindent{\bf Remark.} The proof of the preceding theorem indicates,
at least in the examples considered, that the construction which
to a pair $(\g, U)$ consisting of a semisimple Lie algebra
$\g$  and a finite dimensional $\g$-module associates the closure
of the algebraic
Bethe-Hermite variety $X$ is {\em functorial}: to each homomorphism
$(\g,U)\to (\g',U')$ preserving the Lie algebra and module structures
there corresponds functorially a morphism of algebraic varieties
$X\to X'$. This functor is compatible with the construction of
eigenfunctions
in a sense that should be made more precise.

\section{Weyl group action, Jacobi polynomials, scattering matrices}
\label{sec6}

In this section we  study the Weyl group action on eigenfunctions,
and discuss the relation with multivariable Jacobi polynomials
\cite{HOI}. The
Weyl groups $W$ acts on $U[0]$ (The normalizer $N$ of the Cartan
torus $T=\exp\h$ of the simply connected group with Lie
algebra $\g$ acts on $U$, so $W=N/T$ acts on $U[0]$). Therefore
we have a natural action of $W$ on $U[0]$-valued functions:
$w\in W$ acts as $(w\psi)(\lambda)=w\cdot\psi(w^{-1}\lambda)$.
The Schr\"odinger operator $H$ commutes with this action.

Let $S$ be the set of $\xi\in\h^*$ such that $(\xi,\beta)=(\beta,\beta)$
for some $\beta\in Q_+$.
If $\xi\in \h^*-S$, $E(\xi)$ is isomorphic to $U[0]$ and we have
the explicit expression of the isomorphism in Theorem \ref{first}.

\begin{lemma} Let $\xi\in\h^*-S$.
If $\psi\in E(\xi)$ and $w\in W$,
then $e^{-2\pi i\xi(w^{-1}\lambda)}w\psi$ is
a rational function of $X_1,\dots, X_r$ which is holomorphic
on the complement in $\C^r$ of the root hypersurfaces $X_\alpha=1$,
$\alpha\in \Delta$.
\end{lemma}
\begin{proof} Suppose first that $w=1$. From the explicit expression
\Ref{eqfirst} it is clear that it is a rational function. By Lemma
\ref{regularity},
the poles lie on the root hypersurfaces. Let us assume that
$w=s_k$ is a simple reflection corresponding to the simple root
$\alpha_k$. As root hypersurfaces are permuted under the action
of the Weyl group, it suffices to show
 that $e^{-2\pi i\xi(w^{-1}\lambda)}w\psi$ is
regular at $X_j=0$ for all $j$. This is obvious if $j\neq k$ since
replacing $\lambda$ by $s_k\lambda$ amounts to replacing $X_j$ by
$X_jX_k^{-a_{jk}}$, with $a_{jk}\leq 0$ if $j\neq k$. If $j=k$,
$X_k$ is replaced by $X_k^{-1}$ and we have to check that
the expression in parenthesis in \Ref{eqfirst} is
regular when $X_k\to\infty$, but this follows easily by
counting powers of $X_k$ in the numerator and denominator,
exploiting the fact that $a_j+1\leq p-j+1$. The case of general
$w$ is reduced to this case by writing $w$ as a product of simple
reflections.
\end{proof}

\begin{lemma} Let $\xi\in\h^*$ be generic.
For all $w\in W$, $\psi\mapsto w\psi$ is an isomorphism
from $E(\xi)$ onto $E(w\xi)$.
\end{lemma}

\begin{proof}
Let $\psi\in E(\xi)$. By the previous lemma,
$e^{-2\pi i\xi(w^{-1}\lambda)}w\psi$ is holomorphic at $X=0$.
Also, since $H$ is invariant, $w\psi$ is an eigenfunction
with eigenvalue $4\pi^2(\xi,\xi)$
and therefore $w\psi\in E(w\xi)$.
\end{proof}

\noindent{\it Example.} Let $\g=sl_2$, and $U$ be the
$2s+1$-dimensional irreducible representation. The Weyl
reflection $s_1$ acts as $(-1)^s$ on $U[0]$. Fix
a nonzero $u\in U[0]$ and let $\psi_\xi$ be the eigenfunction
\Ref{gamma}. Then
$(s_1\psi)(\lambda)=(-1)^s\psi_\xi(-\lambda)=
S(\xi)\psi_{-\xi}(\lambda)$, where $S(\xi)$ can be computed
in the limit $X\to 0$. This gives the ``two particle scattering
matrix''
\begin{equation}\label{smatrix}
S(\xi)=(-1)^s\sum_{l=0}^s\frac{(s+l)!\Gamma((\xi,\alpha)-l)}
{l!(s-l)!\Gamma((\xi,\alpha))}
=\prod_{k=1}^{s}\frac{k+(\alpha,\xi)}{k-(\alpha,\xi)}\,.
\end{equation}

We say that a function $\psi$ is Weyl antiinvariant if
$w\psi=\epsilon(w)\psi$ for all $w\in W$.
\begin{proposition}\label{pvan}(cf.\ Prop.\ 3 in \cite{FV})
Let $\psi$ be a meromorphic Weyl antiinvariant solution of the
eigenvalue problem $H\psi=\epsilon\psi$, regular on
\be\h-\cup_{m\in\Z}
\cup_{\alpha\in\Delta}\{\lambda\in\h\,|\,\alpha(\lambda)=m\},\ee
such that $\psi(\lambda+p)=\chi(p)\psi(\lambda)$ for all $p$ in the
 lattice $P^\vee=\{p\in\h\,|\, \alpha(p)\in\Z\}$ and some character
$\chi$ of $P^\vee$,
then $\psi$ extends to a holomorphic function on $\h$.
Moreover, for all $\alpha\in\Delta$ and $m\in\Z$,
\begin{equation}\label{van}
e_\alpha^l\psi
=O((\alpha(\lambda)-m)^{l+1}),
\end{equation}
as $\alpha(\lambda)\to m$.
\end{proposition}

\begin{proof} By periodicity with respect to the coweight lattice $P^\vee$,
we may limit our considerations to the hyperplanes through the
origin.
One proceeds as in Lemma \ref{regularity} by approaching the singular
hyperplane in a transversal direction. The leading term in the
Laurent expansion in the transversal coordinate $x=\alpha(\lambda)$
is
determined by the differential equation
\be
(\alpha,\alpha)\frac{d^2}{dx^2}\psi_0
-{e_\alpha e_{-\alpha}+e_{-\alpha}e_\alpha\over x^2}\psi_0=0.
\ee
Let us decompose $U$ into irreducible representations of
the subalgebra generated by $e_{\pm\alpha}$. Since $\psi$ is
of zero weight, we
may replace $e_\alpha e_{-\alpha}+e_{-\alpha}e_\alpha$ by
\be
C_\alpha=\frac1{(\alpha,\alpha)}h_\alpha h_\alpha
+e_\alpha e_{-\alpha}+e_{-\alpha}e_{\alpha},
\ee
where $h_\alpha=[e_\alpha,e_{-\alpha}]$. If $\psi_0$ belongs
to a $2l+1$ dimensional irreducible representation, i.e., if
$e_\alpha^l\psi\neq 0$ but $e_\alpha^{l+1}\psi= 0$,
then the Casimir element $C_\alpha$
acts as  $(\alpha,\alpha)l(l+1)$. Therefore either $\psi_0\sim x^{l+1}$ or
$\psi_0\sim x^{-l}$. On the other hand, the Weyl reflection
with respect to the hyperplane $\alpha=0$ changes the sign of
$x$ and multiplies the value by $(-1)^l$. Since $\psi$ is
antiinvariant, it follows that
the first possibility is realized and
the function vanishes to order $l+1$.
\end{proof}

In particular, if $\xi$ belongs to the weight lattice $P$,
spanned by the fundamental weights $\omega_1,\dots, \omega_r$,
and $\psi\in E(\xi)$, then the Weyl antiinvariant function
$\psi^W=\sum_{w\in W} \epsilon(w)w\psi$ is a rational
function of $X_{\omega_1},\dots, X_{\omega_r}$
regular on $(\C-\{0\})^r$. Therefore, $\psi$ is a
Laurent polynomial. To apply the previous lemma, $\xi$ should
not be in $S$. This is true if $-\xi$ is dominant, and we
have the following result.

\begin{corollary}
Let $\xi=-\mu$ where $\mu$ is a dominant integral weight and
$\psi\in E(\xi)$. Then $\psi^W=\sum_{w\in W}\epsilon(w)w\psi$ is
a Laurent polynomial in $X_{\omega_1},\dots, X_{\omega_r}$.
\end{corollary}

\noindent{\it Example.} Let $\g=sl_N$, $U=S^{pN}\C^N$, the scalar case, and
let $\omega_1,\dots,\omega_{N-1}$ be the fundamental weights of
$sl_N$.
Let us fix an identification of $U[0]$ with $\C$. Let $\psi_{-\mu}$
be the eigenfunction of the previous corollary, normalized
so that $\psi=e^{-2\pi i\mu(\lambda)}(1+\cdots)$. By the
previous corollary, it is a Laurent polynomial in the $X_{\omega_i}$.
Then the vanishing
property \Ref{van}
 of Proposition \ref{pvan} implies that the Weyl antiinvariant
eigenfunction $\psi^W_{-\mu}$ is divisible in the ring $\C[X_{\omega_i}^{\pm
1}]$
by $\Pi^{p+1}$ where
\be
\Pi=X_{-\rho}\prod_{\alpha\in\Delta_+}(1-X_\alpha)
\ee
The ratio $P=\psi_{-\mu}/\Pi^{p+1}$ is a Weyl invariant polynomial
(since $\Pi$ is antinvariant) and obeys the differential equation
\begin{equation}\label{jac}
-\triangle P+(p+1)2\pi\sum_{\alpha\in\Delta_+}
\cot(\pi\alpha(\lambda))\partial_\alpha P=
\tilde\epsilon P
\end{equation}
($\partial_\alpha$ is the derivative in the direction of $\alpha$),
with $\tilde\epsilon=4\pi^2(\mu+(p+1)\rho,\mu-(p+1)\rho)$.
This equation defines multivariable Jacobi polynomials (or Jack
polynomials) associated to a dominant integral weight $\nu$:
The Jacobi polynomial $P_\nu$ is the unique
solution in $\C[X_{\omega_1}^{\pm 1},\dots,X_{\omega_{N-1}}^{\pm 1}]$
of \Ref{jac}, normalized in such a way that
$P_\nu= X_{-\nu}+\sum_{\beta\in Q_+}a_\beta X_{-\nu+\beta}$
for some coefficients
$a_\alpha\in\C$. The leading term of
 $\psi_{-\mu}/\Pi^{p+1}$ is
$X_{\mu+(p+1)\rho}$. After antisymmetrization, we obtain a leading
term $cX_{w_0(\mu+(p+1)\rho)}$ where $w_0$ is the longest
element of the Weyl group. Thus we obtain a function proportional
to $P_\nu$ by choosing $\mu=w_0\nu-(p+1)\rho$ ($w_0$ is an involution).
Therefore we have the corollary:

\begin{corollary}\label{Jaco}
In the scalar case $\g=sl_N$, $U=S^{pN}\C^N$, with fixed
identification of $U[0]$ with $\C$, let $\nu$ be a
dominant integral weight, and for any $\xi\in\h^*-S$,
let $\psi_{\xi}$ denote the
eigenfunction in $E(\xi)$ given in Theorem \ref{first} with
$u=1$
and $\psi^W_{\xi}=\sum_{w\in W}\epsilon(w)w\psi_\xi$ its antisymmetrization.
Then
\be
P_\nu=c_\nu
\frac{\psi^W_{w_0\nu-(p+1)\rho}}
     {\Pi^{p+1}}\, ,
\ee
for some constant $c_\nu\neq 0$. Here $w_0$ denotes the longest
element of the Weyl group, i.e., the permutation $j\mapsto N+1-j$
of $S_N$.
\end{corollary}
A formula for $c_\nu$ is given below.

We next study more closely the action of the Weyl group.

Let $sl_2(j)$ be the subalgebra of $\g$ generated by
$e_{\pm\alpha_j}$, and for a $\g$-module $U$, let
$U=\oplus_sU_s^{(j)}$ be the decomposition of $U$ viewed as
$sl_2(j)$-module into isotypic components: $U_s^{(j)}$ is
isomorphic to a direct sum of $2s+1$-dimensional
irreducible $sl_2(j)$-modules.

\begin{thm}
Let $\xi$ be generic and denote by $j(\xi)$ the isomorphism
$U[0]\to E(\xi)$ mapping $u$ to the eigenfunction
with leading term $e^{2\pi i\xi(\lambda)}u$.
Then there exists a family of maps
$S(\xi):W\to GL(U[0])$, $w\mapsto S_w(\xi)$ such that
if $\psi=j(\xi)u$, then $w\psi=j(w\xi)S_w(\xi)u$.
These maps have the composition property
\be
S_{w_1w_2}(\xi)=S_{w_1}(w_2\xi)S_{w_2}(\xi).
\ee
 If
$w=s_j$ is a simple reflection then $S_w=S_j$ has
the form
\begin{equation}\label{smat}
S_j(\xi)=P_0^{(j)}+
\sum_{s\geq 1}\prod_{k=1}^{s}\frac{k+(\alpha_j,\xi)}{k-(\alpha_j,\xi)}
P_s^{(j)},
\end{equation}
where $P_s^{(j)}\in\End(U[0])$ is the projection onto $U_s^{(j)}\cap U[0]$.
\end{thm}

\begin{proof}
The first part of this theorem is just a rephrasing of the
preceding lemma. The composition property follows from the definition:
\be
j(w_1w_2\xi)S_{w_1w_2}(\xi)=w_1w_2j(\xi)
=w_1j(w_2\xi)S_{w_2}(\xi)
=j(w_1w_2\xi)S_{w_1}(w_2\xi)S_{w_2}(\xi).
\ee
 If $w=s_j$, we may compute $S_j$ by letting
all $X_l$, $l\neq j$ go to zero. Then the
coefficients $A_L$ in Theorem \ref{first}
 tend to zero except if $L=(j,j,\dots,j)$.
and the formula for $\exp(-2\pi i\xi(\lambda))\psi$
 reduces to an $sl_2$ formula, and we can apply the previous example
in each isotypic component.
\end{proof}

\noindent{\bf Remark.} Note the  similarity between
this scattering matrix and the rational $R$-matrix associated
with general represetations of $sl_2$ \cite{KR}.

\medskip

\noindent {\it Example.} Let $\g=sl_N$, and identify $\h^*$ with
$\C^N/\C(1,\dots,1)$. The Weyl group is the symmetric group
$S_N$ and the simple reflections $s_j$ act on $\C^N$ by
transposition of the $j$th and $j+1$st coordinates. The
corresponding $S_j$ depends only on $(\alpha_j,\xi)=\xi_j-\xi_{j+1}$.
The relations $s_js_{j+1}s_j=s_{j+1}s_js_{j+1}$ translate
into ``unitarity''
\be
S_j(\xi_{j+1}-\xi_j)S_j(\xi_j-\xi_{j+1})=\Id,
\ee
and the Yang--Baxter equation
\be
S_{j}(\xi_{j+1}\!-\!\xi_{j+2})
S_{j+1}(\xi_{j}\!-\!\xi_{j+2})
S_{j}(\xi_{j}\!-\!\xi_{j+1})
\!=\!
S_{j+1}(\xi_{j}\!-\!\xi_{j+1})
S_{j}(\xi_{j}\!-\!\xi_{j+2})
S_{j+1}(\xi_{j+1}\!-\!\xi_{j+2}).
\ee

\begin{corollary}
In the scalar case
($\g=sl_N$, $U=S^{pN}\C^N$)
\be
S_j(\xi)=\prod_{k=1}^{p}\frac{k+(\alpha_j,\xi)}{k-(\alpha_j,\xi)}
\ee
\end{corollary}
\begin{proof} The only isotypic component $U_s^{(j)}$ with non-trivial
intersection with $U[0]$ has $s=p$ in this case.
\end{proof}

Let us conclude with a formula for the constant $c_\nu$ in
Corollary \ref{Jaco}. This constant can be computed from
the leading term in $\psi^W$ which appears in the summand
indexed by $w_0$: $c_\nu^{-1}=\epsilon(w_0)S_{w_0}(w_0(\nu+(p+1)\rho))$.
If we identify $\nu\in\h$ with the diagonal traceless matrix
with diagonal entries $\nu_1,\dots,\nu_N$, a simple calculation
gives
\begin{equation}\label{cnu}
c_\nu^{-1}=\prod_{i>j}\prod_{k=0}^p
\frac
{k+\nu_i-\nu_j-(p+1)(i-j)}
{k-\nu_i+\nu_j+(p+1)(i-j)}\; .
\end{equation}
This constant is clearly different from zero since
for dominant $\nu$ we have $\nu_i<\nu_j$ if $i>j$.

\section{Bethe ansatz in the elliptic case}\label{ellcase}

Let us fix $\tau$ in the upper half-plane, and denote $E_\tau$ the
torus $\C/\Z+\tau\Z$.
We first quote the result of \cite{FV} on eigenfunctions
of the differential operator
\be
H_e=-\triangle+\sum_{\alpha\in\Delta}v(\alpha(\lambda))e_\alpha
e_{-\alpha},
\ee
on $U[0]$-valued functions on $\h$. The elliptic function $v$ is
$v(x)=-\frac
{d^2}{dx^2}\ln\theta(x)$, $\theta(x)=\pi^{-1}\sin(\pi x)
\Pi_1^\infty(1-2q^j\cos(\pi x)+q^{2j})$, $q=e^{2\pi i\tau}$.
It differs from the Weierstrass $\wp$-function by a constant.
\begin{thm}(\cite{FV})
\label{ellBethe}
Suppose $U$ is an irreducible
 highest weight module with highest weight $\Lambda
=\sum_jn_j\alpha_j$ and highest weight vector $v_\Lambda$.
 Set $n=\sum n_j$
and let $c:\{1,\dots,n\}\to\{1,\dots,r\}$
be the unique non-decreasing function such that $c^{-1}\{j\}$ has
$n_j$ elements, for all $j=1$,\dots, $r$.
Then the function parametrized by $t\in\C^n$
\begin{equation}\label{eB66}
\psi(t,\lambda)=
e^{2\pi i\xi(\lambda)}
\sum_{\sigma\in S_n}
w_{\sigma,c}(t,\alpha_1(\lambda),\dots,\alpha_r(\lambda))
f_{c(\sigma(1))}\cdots f_{c(\sigma(n))}v_\Lambda
\end{equation}
(see \Ref{e67} for the definition of $w_{\sigma,c}$)
is an eigenfunction of $H_e$ if the parameters $t_1$, \dots, $t_n$
are a solution of the set of $n$ equations
(``Bethe ansatz equations'')
\begin{equation}\label{eBAE}
\biggl(\sum_{l:l\neq j}
{\theta'(t_j-t_l)\over \theta(t_j-t_l)}\alpha_{c(l)}-
{\theta'(t_j)\over \theta(t_j)}\Lambda
+2\pi i\xi,\alpha_{c(j)}\biggr)=0, \qquad j=1,\dots, n.
\end{equation}
The corresponding eigenvalue $\epsilon$ is
\bea
\epsilon&=&4\pi^2(\xi,\xi)-4\pi i\frac{\partial}{\partial\tau}
 S(t_1,\dots,t_n,\tau), \\
S(t_1,\dots,t_m,\tau)&=&\sum_{i<j}(\alpha_{c(i)},\alpha_{c(j)})
\ln \theta(t_i-t_j)
-\sum_{i}(\Lambda,\alpha_{c(i)})\ln \theta(t_i).
\eea
\end{thm}
\noindent{\bf Remark.} Solutions of the Bethe ansatz equations
are critical points of
\be
\Phi_e(t)=e^{2\pi i(\xi,\sum\alpha_{c(i)}t_i)}\prod_{i<j}
\theta(t_i-t_j)^{(\alpha_{c(i)},\alpha_{c(j)})}
\prod_i\theta(t_i)^{-(\Lambda,\alpha_{c(i)})},
\ee
in the domain $\Phi_e(t)\neq 0$.

As in the trigonometric case, we define the Hermite-Bethe variety by
eliminating the spectral parameter $\xi$ from the Bethe ansatz
equations. The resulting equations are the $n-r$ equations
\begin{equation}\label{HBvar}
\biggl(\sum_{l:l\neq j,j+1}\biggl(
{\theta'(t_j-t_l)\over \theta(t_j-t_l)}-
{\theta'(t_{j+1}-t_l)\over \theta(t_{j+1}-t_l)}\biggr)\alpha_{c(l)}-
\biggl( {\theta'(t_j)\over \theta(t_j)}-
{\theta'(t_{j+1})\over \theta(t_{j+1})}\biggr)\Lambda
,\alpha_{c(j)})\biggr)=0,
\end{equation}
where $j$ runs over the set of indices such that $c(j)=c(j+1)$.

\begin{lemma} The left-hand side of each of the equations \Ref{HBvar}
is a doubly periodic function of each of its arguments $t_i$.
\end{lemma}
This lemma follows easily from the formulas
\be
{\theta'(t+1)\over \theta(t+1)}={\theta'(t)\over \theta(t)},\qquad
{\theta'(t+\tau)\over \theta(t+\tau)}={\theta'(t)\over \theta(t)}-2\pi
i,
\ee
taking into account the zero weight condition $\Lambda=\sum n_j\alpha_j$.

Therefore we may view the equations as algebraic equations on
$E_\tau^n= E_\tau\times\cdots\times E_\tau$, a product of elliptic curves.
Moreover, we have an action of $S_c=S_{n_1}\times\cdots\times S_{n_r}$
permuting $t_i$ with the same $c(i)$, that maps solutions to solutions,
and does not change $\psi$. Let $D(c)$ be the set of $t\in E_\tau^n$
such that $t_i=t_j$ for some $i\neq j$ with
$(\alpha_{c(i)},\alpha_{c(j)})
\neq 0$, or $t_i=0$ for some $i$ with $(\Lambda,\alpha_{c(i)})\neq 0$.
On this set the Bethe ansatz equations are singular.

\begin{definition}
With the notation of the Theorem, assume that $n_j>0$ for all $j$.
The Hermite--Bethe variety
$H\!B(c)$ is the subvariety of $(E_\tau^n-D(c))/S_c$ defined by the
equations \Ref{HBvar}.
\end{definition}

The remaining $r$ equations determine $\xi$ as a function of
a solution $t$ of \Ref{HBvar}. They can be chosen to be the
equations \Ref{eBAE} with $j=n_1,n_1+n_2,\dots,n$.
We see from this formula that if $t_j$ is replaced by $t_j+n+m\tau$,
then
$\xi$ is shifted by $-m\alpha_{c(j)}$. It is easy to see that these
replacements
do not change the eigenfunction $\psi$. Therefore we have
a map $\xi:H\!B(c)\to \h^*/Q$ mapping $t$ to
$\xi$, and $H\!B(c)$ parametrizes eigenfunctions $\psi$ such that
$\psi(\lambda+\omega)=e^{2\pi i\xi(\omega)}\psi(\lambda)$,
$\omega\in P^\vee$.

One would like to prove that ``all'' eigenfunctions are obtained in this
way. In the general case not much is known, however in the scalar
case and the case of the adjoint representation we have the
following result:

\begin{thm}\label{eCBS} Let $\g=sl_N$, $U=S^{pN}\C^N$ or the adjoint
representation. Then
for each generic $\xi\in\h$ there  are $\dim(U[0])$ solutions
$t\in\C^n$ of the Bethe ansatz equations \Ref{eBAE}.
The corresponding eigenfunctions obey
$\psi(\lambda+\omega)=e^{2\pi i\xi(\omega)}\psi(\lambda)$,
$\omega\in P^\vee$, and are linearly independent.
\end{thm}

The proof of this theorem is essentially the same as in the trigonometric
case (Theorem \ref{cBS}): in the case of $sl_2$ one has Hermite's
result, and one proceeds by induction in $N$. The point is that one only
has to use the asymptotic behavior of the theta functions involved,
and this is the same as in the trigonometric case.

\section{The $q$-deformed case}\label{quantum}

We give here the $q$-deformed version of our first formula.
Our conventions for quantum groups are the following. Let
$q$ be a complex number different from 0, 1 or -1. We fix a
logarithm of $q$, so that $q^a$ is defined for all $a\in\C$.
We normalize
the invariant bilinear form on $\g$ in such a way that $(\alpha_j,
\alpha_l)\in\Z$, for all $j,l\in\{1,\dots,r\}$.
The Drinfeld--Jimbo
 quantum universal enveloping algebra $U_q\g$, a Hopf algebra,
 is the algebra with unit
over
$\C$ generated by $f_j$, $e_j$, commuting
elements $k_j$, and their inverses $k_j^{-1}$, $j=1,\dots, r$
with relations $k_je_lk_j^{-1}=q^{(\alpha_j,\alpha_l)}e_l$,
$k_jf_lk_j^{-1}=q^{-(\alpha_j,\alpha_l)}f_l$, $e_jf_l-f_le_j=
\delta_{jl}(k_j-k_j^{-1})/(q-q^{-1})$, and deformed Serre
relations $s_a(q)=0$, $a=1,\dots,2m$, (see \cite{CP}).
 The coproduct is defined on generators
to be $\Delta(f_j)=f_j\otimes 1+k^{-1}_j\otimes f_j$,
$\Delta(e_j)=e_j\otimes k_j+1\otimes e_j$ and
$\Delta(k_j^{\pm1})=k_j^{\pm1}\otimes k_j^{\pm1}$. We consider
modules $M$ over $U_q\g$ admitting a weight decomposition
into finite dimensional weight spaces $M[\mu]$, $\mu\in\h^*$,
on which $k_j$ acts as $q^{(\alpha_j,\mu)}$.
In particular, the Verma module of weight $\mu$
is the quotient $M_\mu=U_q\g/I(\mu)$ by the left ideal $I(\mu)$
generated by $k_j-q^{(\alpha_j,\mu)}1$, $e_j$, $j=1,\dots,r$. It
is generated by its highest weight vector $v_\mu$, the class of 1.

If $U$ is a finite dimensional $U_q\g$-module, and
$\Phi$ a homomorphism of $U_q\g$-modules
$\Phi: M_{\xi-\rho}\to M_{\xi-\rho}\otimes U$, we may define, as
in the classical case,
\begin{equation}\label{psi}
\psi(\lambda)=
{\sum_\mu e^{2\pi i\mu(\lambda)}\tr_{M_{\xi-\rho}[\mu]}\Phi
\over
\sum_\mu e^{2\pi i\mu(\lambda)}\tr_{M_{-\rho}[\mu]}1}
\in e^{2\pi i\xi(\lambda)}\C[[X_1,\dots,X_r]].
\end{equation}
Recall that $X_j=e^{-2\pi i\alpha_j(\lambda)}$.
As in the classical case, we have for generic $\xi$, $q$, and
for each $u$ in the zero-weight space $U[0]$ of a finite
dimensional $U_q\g$-module $U$, a
unique homomorphism of $U_q\g$-modules
$\Phi: M_{\xi-\rho}\to M_{\xi-\rho}\otimes U$, sending the
generating vector $v_{\xi-\rho}$ to $v_{\xi-\rho}\otimes u$ \cite{EK2}.
So, in the generic case, the $\psi$-function is uniquely determined
by $q,\xi$ and a vector $u\in U[0]$. It is a formal power series,
 but its explicit expression below shows that it is actually a
meromorphic function of $\lambda$.

If $\g=sl_N$ and $U$ is a deformation of the $pN$th symmetric
power of $\C^N$, it was shown by Etingof and Kirillov
\cite{EK2} that $\psi(\lambda)$ is
a common eigenfunction of a commuting family of difference
operators (related by conjugation by a known function
to the $A_{N-1}$-Macdonald operators).

As in the classial case, the image of the generating vector is a singular
vector (a vector killed by $e_i$, $i=1,\dots,r$) of weight $\xi-\rho$
 and all singular
vectors of weight $\xi-\rho$ correspond to some homomorphism.
 Set
$f_L=f_{l_1}\cdots f_{l_m}$ for a multiindex $L=(l_1.\dots,l_m)$.
Our formula gives, for each $u\in U[0]$,
$\psi$ in terms of the coefficients $u_L$ in the formula of
the the unique singular vector of the form
$v\otimes u+\sum_Lf_Lv_{\xi-\rho}\otimes u_L$
These coefficients
are given explicitly in terms of $u$ and the inverse Shapovalov
matrix, see \cite{EK3}.
\begin{thm}\label{last}
 Let $\xi\in\h^*$ be generic and let
 $v_{\xi-\rho}\otimes u+\sum_Lf_Lv_{\xi-\rho}\otimes u_L$
be a singular vector of weight $\xi-\rho$
in $M_{\xi-\rho}\otimes U$. Let $\Phi\in\Hom_{U_q\g}(M_{\xi-\rho},
M_{\xi-\rho}\otimes U)$ be the corresponding homomorphism,
and $\psi$ the $\psi$-function \Ref{psi}. Then
$\psi(\lambda)=
e^{2\pi i\xi(\lambda)}(u+\sum_LA_L(\lambda)u_L)$, with
\be
A_{(l_1,\dots,l_p)}(\lambda)=\sum_{\sigma\in S_p}
q^{a(L,\sigma)}
\left(\prod_{j=1}^p
\frac
{X_{l_{\sigma(j)}}^{a_j(\sigma)+1}}
{1-X_{l_{\sigma(1)}}\cdots X_{l_{\sigma(j)}}
q^{|\alpha_{l_{\sigma(1)}}+\cdots+\alpha_{l_{\sigma(j)}}|^2}}\right)
f_{l_{\sigma(1)}}\cdots f_{l_{\sigma(p)}}
\ee
where $a_j(\sigma)$ is the cardinality of the set of $m\in\{j,\dots,p-1\}$
such that $\sigma(m+1)<\sigma(m)$, $X_j=\exp(-2\pi i\alpha_j(\lambda))$,
 and
\bea
a(L,\sigma)&=&\sum_{j=1}^p(\alpha_{l_j},\rho-\xi)
+\sum_{k<j,\sigma(k)>\sigma(j)}
(\alpha_{l_{\sigma(j)}},\alpha_{l_{\sigma(k)}})
+\sum_{m\in S}(\sum_{j=1}^m\alpha_{l_{\sigma(j)}},
\sum_{j=1}^m\alpha_{l_{\sigma(j)}}),\\
S&=&\{m\in\{1,\dots,p-1\}|\sigma(m)>\sigma(m+1)\}\cup\{p\}.
\eea
\end{thm}

We conclude this section by giving a conjectural formula for Macdonald
polynomials, which is a $q$-deformation of Corollary \ref{Jaco}. The
$A_{N-1}$ Macdonald polynomials \cite {Mac} are symmetric polynomials
$P_\nu(x,q,t)$ (we use the notation of \cite{EK2}) in $N$
variables $x_1$, \dots $x_N$, depending on parameters $q$ and $t$. They
are labeled by  dominant integral weights $\nu$ of $gl_N$, i.e., decreasing
sequence $\nu_1\geq\dots\geq\nu_N$ of non-negative integers. We
consider the case where $t=q^k$ for some positive integer $k$ as in
\cite{EK2}.  The symmetric  polynomials $P(x,q,q^k)$ are uniquely
characterized by having leading term $x_1^{\nu_1}\cdots x_N^{\nu_N}$
(as $x_i/x_{i+1}\to\infty$,  $i=1,\dots,N-1$)
with unit coefficient and by being orthogonal with respect to the
inner product $(f,g)=$ constant term of $f(x_1,\dots, x_N)g(x_1^{-1},
\dots x_N^{-1})\Delta(x)$,
where
\be
\Delta(x)=\prod_{i\neq j}\prod_{m=0}^k(1-q^{2m}x_i/x_j).
\ee
Obviously $P_{\nu+(1,\dots,1)}=x_1\dots x_NP_\nu$ so it is sufficient
to evaluate Macdonald polynomials on the hypersurface $x_1\cdots x_N=1$.
Then $\nu$ may be condidered modulo $\Z(1,\dots,1)$, i.e., as
$sl_N$ dominant weight.
The $q$-analogue of Corollary \ref{Jaco} is the following conjecture,
which can be proved with the same method as in the classical case
if $N=2$ or 3, but is open in the general case.

\begin{conjecture}\label{qJaco}
Consider the scalar case $\g=sl_N$, with $\h=\{\lambda\in\C^N\,|
\,\sum\lambda_i=0\}$, $U=S^{pN}\C^N$, and fix an identification
of $U[0]$ with $\C$. Set $x_j=e^{2\pi i\lambda_j}$.  Let $\nu\in \h^*$
 be a
dominant integral weight, and for  $\xi\in\h^*$,
let $\psi_{\xi}$ denote the
function   given in Theorem \ref{last} with
$u=1$
Then
\be
P_\nu(x,q,q^{p+1})=c_\nu\sum_{w\in S_N}
\frac{\psi_{w_0\nu-(p+1)\rho}(w\lambda)}{\delta(w\lambda)},
\ee
where
\be
\delta(\lambda)=
     \prod_{m=0}^p\prod_{j<l}(e^{\pi i(\lambda_j-\lambda_l)}
-q^{-2m}e^{\pi i(\lambda_l-\lambda_j)}),
\ee
for some constant $c_\nu\neq 0$. Here $w_0$ denotes the longest
element of the Weyl group, i.e., the permutation $j\mapsto N+1-j$
of $S_N$.
\end{conjecture}
As before, the constant $c_\nu$ may be computed in terms of
the $q$-analogue of the scattering matrix \Ref{cnu}
\be
c_\nu^{-1}=\prod_{i>j}\prod_{m=0}^p
\frac
{[m+\nu_i-\nu_j-(p+1)(i-j)]_q}
{[m-\nu_i+\nu_j+(p+1)(i-j)]_q}\; , \quad [x]_q=\frac{q^x-q^{-x}}{q-q^{-1}}\,.
\ee

\section{Proof of Theorems \ref{first} and \ref{last}}\label{proofs}

We give the proof of Theorem \ref{last}. The proof of Theorem
\ref{first} is essentially a special case.

The proof is in two parts. We first prove that the calculation
can be done in the algebra without Serre relation
(Prop.\ \ref{free}), and
then do this calculation, reducing it to a combinatorial problem.

We denote by $U_q\bor$ the subalgebra  of $U_q\g$
generated by $f_j,k^{\pm 1}_j$,
$j=1,\dots,r$ and
by $U_q\n$ the subalgebra generated by $f_j$, $j=1,\dots,r$.
We have $\Delta(U_q\n)\subset U_q\bor\otimes U_q\n$.
Let us introduce a $Q$-grading of these algebras by setting
$\deg(f_j)=\alpha_j$, $\deg(k_j)=0$.
What we need to know about the deformed Serre relations is that
(i) $U_q\n$ is the quotient of the free algebra
 generated by $f_1,\dots,f_r$ by the ideal $J_q$ generated by
the deformed Serre relations $s_1(q)$,\dots,$s_m(q)$, (ii)
$s_1(q)$, \dots $s_m(q)$ are homogeneous polynomials in the $f_j$ with
coefficients in $\Z[q,q^{-1}]$ reducing to the
(classical) Serre relations at $q=1$, and (iii)
$\Delta(s_a(q))=s_a(q)\otimes 1+K\otimes s_a(q)$ for some
$K=\prod k_j^{-m_j}$.

We will also need the fact that the dimensions of the spaces
of fixed degree in $U_q\n$ are independent of $q$ and coincide
with the dimensions of the classical enveloping algebra
$U\n$ of the Lie subalgebra $\n\subset\g$ generated by
$f_j=e_{-\alpha_j}$, $j=1,\dots,r$ (see, e.g., \cite{CP}).

The first observation is that the computation of the trace,
given the components of the singular vector is a computation
in $U_q\n$:

\begin{lemma}\label{le00}
Let for $w\in M_{\xi-\rho}$,
$\Phi^\n(w): M_{\xi-\rho}\to
M_{\xi-\rho}\otimes U_q\n$
 be the unique
homomorphism of $U_q\n$-modules mapping $v_{\xi-\rho}$
to $w\otimes 1$. Then, in the notation of Theorem \ref{last},
\be
A_L(\lambda)=
{\sum_\mu e^{2\pi i\mu(\lambda)}\tr_{M_{\xi-\rho}[\mu]}
\Phi^\n(f_Lv_{\xi-\rho})
\over
\sum_\mu e^{2\pi i\mu(\lambda)}\tr_{M_{-\rho}[\mu]}1}
\ee
\end{lemma}
\noindent
This lemma follows immediately from the fact that
Verma modules are free over $U_q\n$.

Let us introduce algebras without Serre relations.

\begin{definition} Let  $U_q\tilde\bor$ be the algebra over $\C$
generated by $f_1,\dots,f_r,k^{\pm 1}_1,\dots,k^{\pm 1}_r$ with relations
$k_jf_kk_j^{-1}=q^{-(\alpha_j,\alpha_k)}f_k$, $k_jk_l=k_lk_j$,
 $k_jk_j^{-1}=1$ and with coproduct
as in $U_q\g$. Let $U_q\tilde\n\subset
U_q\tilde\bor$ be the free algebra on generators $f_1,\dots, f_r$.
The {\em Verma module} $\tilde M_\mu$ over $U_q\tilde\bor$
of highest weight $\mu\in\h^*$ is the left module $U_q\tilde\bor/I(\mu)$
where $I(\mu)$ is the left ideal generated by $k_j-q^{(\mu,\alpha_j)}1$,
$j=1,\dots,r$.
The image of $1$ in $\tilde M_\mu$ is denoted by $\tilde v_\mu$.
\end{definition}

\noindent
The Verma module $\tilde M_\mu$ has a weight decomposition with
finite dimensional weight spaces, and is freely generated
by $\tilde v_\mu$ as a
$U_q\tilde\n$-module. In particular, $M_{\mu}=\tilde M_\mu/J_q\tilde
v_\mu$.
Moreover,
$\Delta(U_q\tilde\n)\subset U_q\tilde\bor\otimes U_q\tilde\n$.
Therefore the construction of the
homomorphisms $\Phi^\n(\tilde w):\tilde M_{\xi-\rho}
\to\tilde M_{\xi-\rho}\otimes U_q\tilde\n$ works
also in this case and traces make sense (as a formal power
series).
\newcommand{\serre}{{\frak s}}

\begin{proposition}\label{free} Let for $w\in M_{\xi-\rho}$,
$\Phi^\n(w): M_{\xi-\rho}\to
M_{\xi-\rho}\otimes U_q\n$
 be the unique
homomorphism of $U_q\n$-modules mapping $v_{\xi-\rho}$
to $w\otimes 1$, and for $\tilde w\in\tilde M_{\xi-\rho}$
let $\Phi^\n({\tilde w})$ be the same object for the algebra
without Serre relations. Then for any $\tilde w$ projecting
to $w$ under the canonical projection $\tilde M_{\xi-\rho}
\to M_{\xi-\rho}$,
\begin{equation}\label{traces}
{\sum_\mu e^{2\pi i\mu(\lambda)}\tr_{M_{\xi-\rho}[\mu]}
\Phi^\n(w)
\over
 \sum_\mu e^{2\pi i\mu(\lambda)}\tr_{M_{-\rho}[\mu]}1}
={\sum_\mu e^{2\pi i\mu(\lambda)}\tr_{\tilde M_{\xi-\rho}[\mu]}
\Phi^\n(\tilde w)
\over
  \sum_\mu e^{2\pi i\mu(\lambda)}\tr_{\tilde M_{-\rho}[\mu]}1}
\end{equation}
\end{proposition}
\noindent The proof of this proposition requires some preparation,
and will be completed after Lemma \ref{le4} below.
\begin{lemma}\label{le0}
If $x\in U_q\n$ is homogeneous of degree $\alpha=\sum_jm_j\alpha_j\in Q_+$,
then
$\Delta(x)$ can be written as
$\Delta(x)=\prod_jk_j^{-m_j}\otimes x+\sum x'_j\otimes x''_j$,
where $x''_j$ are homogeneous of degree $<\alpha$.
\end{lemma}
\noindent
This lemma is an easy consequence of the form of the coproduct
of the generators $f_j$.

\begin{definition} Let $\tilde\n$ be the free Lie algebra on
$r$ generators
$f_1,\dots, f_r$ (see \cite{Hu}). An {\em iterated bracket of length one}
is an element $f_j$ of this set. An {\em iterated bracket of
length $l> 1$} is defined recursively as an expression of
the form $[a,b]$, where $a$, and $b$ are iterated brackets of length
$l_1$, $l-l_1$ respectively, for some $l_1\in\{1,\dots, l-1\}$.
An iterated bracket of length $l$ is called {\em simple} if
it is of the form $[f_{i_1},[f_{i_2},[\cdots,f_{i_l}]]]$.
\end{definition}
\begin{lemma}\label{le1}
Let $a,b\in\tilde\n$, and $a$ be an iterated bracket. Then $[a,b]$ is a linear
combination of elements of the form
\begin{equation}\label{form}
[f_{i_1},[f_{i_2},[\cdots,[f_{i_l},b]]]].
\end{equation}
\end{lemma}
\begin{proof}
We use induction in the length of $a$. If the length of $a$ is
one there is nothing to prove. If $a=[a_1,a_2]$ is of length $l$,
with $a_i$ of length $l_i<l$, the Jacobi identity
gives $[a,b]=[a_1,[a_2,b]]-[a_2,[a_1,b]]$. By the induction hypothesis,
$b'=[a_1,b]$ and $b''=[a_2,b]$ are a linear combination of elements
of the desired form \Ref{form}. Then we use the induction hypothesis
once more with $b$ replaced by $b'$ and $b''$.
\end{proof}

\noindent
We apply this lemma to the following situation. The Lie
algebra $\n\subset\g$  is the quotient of the free Lie algebra $\tilde\n$
on $r$ generators by the Lie ideal $\serre$ generated by
Serre relations $s_1$, \dots, $s_m$, which are simple iterated brackets
in $\tilde\n$.

\begin{lemma}\label{le2} Denote by $\tilde\n$  the free Lie algebra on
$r$ generators $f_1, \dots, f_r$, and let $s_1,\dots, s_m$ be simple
iterated brackets in $\tilde\n$.   Let $\serre$ be the Lie
 ideal generated by $s_1,\dots, s_m$, and assume that
$d=\dim\tilde(\n/\serre)<\infty$.  Then there exists a basis $b_1, b_2,
\dots$ of $\tilde \n$ consisting of simple iterated brackets such that
$b_{d+1}, b_{d+2},\dots$ is a basis of $\serre$, and such that, for $j>d$,
$b_j$ is of the form
\be
[f_{i_1},[f_{i_2},[\cdots,[f_{i_l},s_k]]]].
\ee
\end{lemma}

\begin{proof}
By definition every element in $\tilde\n$ is a linear combination
of iterated brackets. The ideal $\serre$ is  spanned by iterated
brackets containing at least one of the $s_j$.
Using the skew-symmetry of the bracket, we see that every
element of $\serre$ is a linear combination of elements of the
form $[a_1,[a_2,[\dots,s_k]]]$ for some iterated brackets
$a_j$. The claim then follows  from the preceding lemma.
\end{proof}

\noindent
By the Poincar\'e--Birkhoff--Witt theorem, the universal
enveloping algebra $U\tilde\n$  has a basis
$b^J=b_k^{j_k}\cdots b_2^{j_2}b_1^{j_1}$ labeled by the set $\cal J$ of
sequences $J=(j_1, j_2,\dots)$ of non-negative integers
with $j_l=0$ for all sufficiently large $l$. As the Verma module
$\tilde M_\mu$ is freely generated by the highest weight vector
$\tilde v_\mu$ as a
$U\tilde\n$-module, $(b^J\tilde
v_\mu)_{J\in\cal J}$ is a basis of $\tilde M_\mu$.

Let us extend this to the $q$-deformed case. Note that
$U_q\n$ as an algebra is the same for all $q$.

\begin{lemma}\label{le3}
There exists a sequence $b_1(q),b_2(q),\dots$ of homogeneous elements in
$U\tilde\n\otimes\C[q,q^{-1}]$ such that
\begin{enumerate}
\item[(i)] $b_j(q)=b_j \mod (q-1)\C[q,q^{-1}]$
\item[(ii)] If $q$ is a  transcendental number,
the elements $b^J(q)=\cdots b_2(q)^{j_2}b_1(q)^{j_1}$
form a basis of $U_q\tilde \n$.
\item[(iii)] For $j\geq d+1$, $\Delta(b_j(q))=b_j(q)\otimes 1
\mod U_q\tilde \bor\otimes J_q$.
\end{enumerate}
\end{lemma}

\begin{proof}
We may take $b_j(q)=b_j$ if $j\leq d$. If $a\in U_q\tilde\n$ is a homogeneous
element of degree
 $\beta\in Q$, set $\ad_q(f_j)a
=f_ja-q^{(\beta,\alpha_j)}af_j$. If $j>d$ and
$b_j=\ad(f_{j_1})\cdots \ad(f_{j_l})s_k$, set
\be
b_j(q)=\ad_q(f_{j_1})\cdots \ad_q(f_{j_l})s_k(q).
\ee
It is clear that (i) holds. (ii) holds too, since the determinant
of the matrix expressing, in each homogeneous component of $U\tilde n\otimes
\C[q,q^{-1}]$,
$b^J(q)$ in terms of $b^L$ has a determinant in $\Q[q,q^{-1}]$ with
the value 1 at $q=1$, and is therefore invertible if $q$ is transcendental.
 As for (iii), we know that
$s_k(q)$ has the required property since $\Delta(s_k(q))
=s_k(q)\otimes 1+K\otimes s_k(q)$ for some $K$. Moreover
if $a\in U_q\n$ of degree $\beta$
 obeys $\Delta(a)=a\otimes 1\mod U_q\bor\otimes J_q$,
then
\bea
\Delta(\ad_q(f_j)a)&=& f_ja\otimes 1+k_j^{-1}a\otimes f_j
-q^{(\beta,\alpha_j)}(af_j\otimes 1+ak_j^{-1}\otimes f_j)
\mod U_q\tilde\bor\otimes J_q
\\
&=&
\ad_q(f_j)a\otimes 1\mod U_q\tilde\bor\otimes J_q,
\eea
since $k_j^{-1}a=q^{(\beta,\alpha_j)}ak^{-1}_j$.
Therefore, by induction, $\Delta b_j(q)=b_j(q)\otimes 1
\mod U_q\tilde b\otimes J_q$.
\end{proof}

\noindent
Suppose that $q$ is transcendental, and fix a sequence
$b_j(q)$, as in Lemma \ref{le3}
and thus a basis ($b^J(q)$) of $U_q\tilde\n$. To simplify the notation,
we will write, for any $J\in\cal J$, $\deg(J)=\deg(b^J(q))$. Let
$\cal J''$,  be the set of $J=(j_1,j_2,\dots)\in\cal J$ such that
$j_l=0$ if $l>d$, and $\cal J'$ be the set of $J$ such that $j_l=0$
if $l\leq d$. Then $\cal J=\cal J'\times\cal J''$ canonically, and
we may write the basis as $b^{J'}(q)b^{J''}(q)$, $(J',J'')
\in\cal J'\times \cal J''$. Then $b^{J'}(q)b^{J''}(q)\in J_q$
if $J'$ is non-trivial. Since the dimensions of the homogeneous
components of $U_q\n$ are the same as in the classical case,
the basis elements with $J'$ non-trivial form a basis of
$J_q$ and the classes of $b^{J''}(q)$
form a basis of $U_q\n=U_q\tilde\n/J_q$.
\begin{lemma}\label{le4}
Suppose that $q\in \C$ is transcendental.
Then the  elements $b^J(q)$, ${J\in\cal J'}$
form a basis of the subalgebra of $U_q\tilde\n$ consisting
of elements $x$ such that $\Delta(x)=x\otimes 1\mod
U_q\tilde\bor\otimes J_q$
\end{lemma}
\begin{proof}
Call $B$ this subalgebra.
By Lemma \ref{le3}, (iii), the linearly independent
elements $b^J(q)$, $J\in\cal J'$
belong to $B$. What is left to prove is that any $x\in B$
can be written as a linear combination of these elements.
Write $x=\sum_{\cal J'\times\cal J''}a_{J',J''}b^{J'}(q)b^{J''}(q)$,
with complex coefficients $a_{J',J''}$. Let $Z\subset Q_+$ be the set
of $\beta$ such that there exist $J',J''$ with deg$(J'')=\beta$ and
$a_{J',J''}\neq 0$. An element $\beta\in Z$ is called maximal
if $\beta'>\beta$ implies $\beta'\not\in Z$. Obviously,
every element in $Z$ is $\leq$ some maximal element. Therefore,
if we show that the only maximal element is $0$, we have
proved the lemma.

By Lemma \ref{le3},
\be
\Delta(x)=\sum a_{J',J''}(b^{J'}(q)\otimes 1)\Delta(b^{J''}(q))
\mod U_q\tilde\bor\otimes J_q.
\ee
Let $\beta$ be maximal. The terms whose second factor
has degree $\beta$ are, by Lemma \ref{le0},
\be
\Delta(x)=\cdots+\sum_{\deg(J'')=\beta}a_{J',J''}
b^{J'}(q)\otimes b^{J''}(q)+\cdots \mod U_q\tilde\bor\otimes J_q.
\ee
These terms must vanish if $\beta\neq 0$ since $x\in B$.
But the elements $b^{J''}$ are linearly independent, even mod
$J_q$. And, by construction, $a_{J',J''}\neq 0$
for some $J',J''$ with deg$(J'')=\beta$. Therefore $\beta=0$.
\end{proof}

\noindent
The proof of Proposition \ref{free} is a calculation of the traces
using the basis $b^J(q)\tilde v_{\xi-\rho}$, ${J\in\cal J}$
of $\tilde M_{\xi-\rho}$
and $b^J(q)v_{\xi-\rho}$, ${J\in\cal J''}$ of $M_{\xi-\rho}$.
It is sufficient to prove Proposition free in the case where
$q$ is transcendental, since all traces over weight spaces are
clearly rational functions of $q$.
Let $\tilde w$ and $w$ be as in Proposition \ref{free}.
If $J\in \cal J''$,
$\Delta b^{J}(q)\tilde w=\sum_{L\in\cal J}
 b^{L}(q)\otimes A_{JL}$, for some $A_{JL}\in U_q\tilde n$.

The numerator of the left-hand side of
\Ref{traces} is then
\be\sum_\mu e^{2\pi i\mu(\lambda)}\tr_{M_{\xi-\rho}[\mu]}
\Phi^\n(w)
=\sum_{J\in \cal J''} A_{JJ}
X_{\xi-\rho+{\deg}(J)}\mod J_q.\ee
On the other hand, the calculation of the numerator on the right-hand
side of \Ref{traces}
involves (see Lemma \ref{le3}, (iii))
\be
\Delta(b^{J'}(q)b^{J''}(q))w\otimes 1=
\sum_{L\in \cal J}b^{J'}(q)b^L(q)\otimes A_{J''L}\mod U_q\bor\otimes J_q.
\ee
Now we claim that the only terms in this summation that may give
a non-trivial contribution to the trace are those with $L\in \cal J''$.
Let indeed $b^L(q)=b^{L'}(q)b^{L''}(q)$, with $L'\in\cal J'$ and
$L''\in\cal J''$. Since both $b^{J'}(q)$ and $b^{L'}(q)$ are
in the algebra defined in Lemma \ref{le4}, their product is
also in this algebra and is thus a linear combination of
$b^{M'}(q)$, $M'\in \cal J'$, deg$(M')=\deg(J')+\deg(L')$. If
$L'\neq 0$, this degree is strictly larger than deg($J'$) and
therefore does not contribute to the trace.
The trace is therefore
\bea
\sum_\mu e^{2\pi i\mu(\lambda)}\tr_{\tilde M_{\xi-\rho}[\mu]}
\Phi^\n(\tilde w)
&=&
\sum_{J'\in\cal J'} \sum_{J''\in\cal J''} A_{J''J''} X_{\rho-\xi+\deg
(J'')} X_{\deg(J')}\mod J_q\\
&=&
\bigl(\sum_{J'\in\cal J'} X_{\deg(J')}\bigr)
\sum_\mu e^{2\pi i\mu(\lambda)}\tr_{M_{\xi-\rho}[\mu]}
\Phi^\n(w) \mod J_q.
\eea
On the other hand,
\bea
  \sum_\mu e^{2\pi i\mu(\lambda)}\tr_{\tilde M_{-\rho}[\mu]}1
&=&
\sum_{J'\in\cal J'} \sum_{J''\in\cal J''} X_{\rho+\deg(J'')+\deg(J')}
\\
&=&
\bigl(\sum_{J'\in\cal J'} X_{\deg(J')}\bigr)
 \sum_\mu e^{2\pi i\mu(\lambda)}\tr_{M_{-\rho}[\mu]}1.
\eea
Taking the ratio of the last two expression completes the
 proof of  Proposition
\ref{free}.

We now turn to the actual calculation of the ratio of traces
on the right-hand side of \Ref{traces}.
 We use the basis
$(f_J=f_{j_1}\dots f_{j_m})$ of $U_q\tilde\n$, labeled by
finite sequences $J$ in $\{1,\dots,r\}$ of arbitrary length
$m\geq 0$ (if $m=0$, we set $f^{(\ )}=1$), and we may assume that
$\tilde w=f_L\tilde v_{\xi-\rho}$, with $L=(l_1,\dots,l_p)$.
By definition, the trace in the numerator of \Ref{traces}
is the sum over $J=(j_1,\dots,j_m)$ of the diagonal
elements $B_{JJ}$ in
\begin{equation}\label{qt1}
\Delta(f_J)f_L\tilde v_{\xi-\rho}\otimes 1
=\sum_Mf_M\tilde v_{\xi-\rho}\otimes B_{MJ},
\end{equation}
weighted by a factor $X_{\rho-\xi}\prod_{k=1}^{m} X_{j_k}$.
 Expanding the coproduct on the left-hand side of \Ref{qt1}
 gives $2^m$ terms, each of them of the
form
\be
q^{a}f_{j_{b_1}}\cdots f_{j_{b_s}}f_L\tilde v_{\xi-\rho}
\otimes f_{j_{d_1}}\cdots
f_{j_{d_{m-s}}},
\ee
labeled by subsets $B=\{b_1<\cdots< b_s\}$ of $\{1,\dots,m\}$
with complement $D=\{d_1<\cdots<d_{m-s}\}$. The exponent of
$q$ is
\begin{equation}\label{defa}
a=(\xi-\rho+\sum_j\alpha_{l_j},\sum_{d\in D}\alpha_{j_d})
+\sum_{B\ni b>d\in D}(\alpha_{j_b},\alpha_{j_d}).
\end{equation}
This term gives a contribution to the trace if and only
if
\begin{equation}\label{qt2}
(j_{b_1},\dots,j_{b_s},l_1,\dots, l_p)=(j_1,\dots,j_m).
\end{equation}
If it does, the contribution is
\be
q^a
X_{\xi-\rho}X_{j_1}\cdots X_{j_m}f_{j_{d_1}}\cdots f_{j_{d_{m-s}}}.
\ee
Note that the $j_{d_i}$ are necessarily equal to the $l_i$
up to ordering. In particular $m-s=p$.
The terms contributing to the trace in the numerator are
therefore in one-to-one correspondence with pairs consisting
of a finite sequence
$(j_1,\dots,j_m)$ and a subset $B\subset\{1,\dots,m\}$ obeying
\Ref{qt2}. Let  $B=\{1\leq b_1<b_2<\cdots<b_s\}\subset\N$ be given.
If $b_j=j$ for all $j$, set $t=s+1$. Otherwise let
$t$ be the smallest integer so that $b_t>t$. Then
 the condition \Ref{qt2} on $J$ determines
uniquely $j_t,j_{t+1},\dots,j_m$ in terms of $L$, and gives no
constraint on $j_1,\dots,j_{t-1}$. Therefore we can restrict
our attention to the case $t=1$: the general solution $(J,B)$ of
\Ref{qt2} is
$J=(j_1,\dots,j_{t-1},j'_1,\dots,j'_m)$, $B=\{1,\dots,t-1,
b'_1+t-1,\dots,b'_s+t-1\}$ where $(J',B')$ is a solution with
$b'_1>1$, and $j_1,\dots,j_{t-1}$ are arbitrary. The contribution
to the trace of such a solution $(J,B)$ is $X_{j_1}\cdots X_{j_{t-1}}$
times the contribution of $(J',B')$. The sum over all such solutions
with fixed $(J',B')$ gives a factor that cancels with the
trace in the denominator (except for $X_\rho$).
\be
  \sum_\mu e^{2\pi i\mu(\lambda)}\tr_{\tilde M_{-\rho}[\mu]}1
=X_{\rho}\sum_{m=0}^\infty
\sum_{j_1,\dots,j_{m}}X_{j_1}\cdots X_{j_m}
\ee
 This implies
\begin{lemma}\label{le5}
Let $L=(l_1,\dots,l_p)$.
\begin{equation}\label{qt3}
{\sum_\mu e^{2\pi i\mu(\lambda)}\tr_{\tilde M_{\xi-\rho}[\mu]}
\Phi^\n(f_L\tilde v_{\xi-\rho})
\over
  \sum_\mu e^{2\pi i\mu(\lambda)}\tr_{\tilde M_{-\rho}[\mu]}1}
=e^{2\pi i\xi(\lambda)}\sum_{(J,B)}q^aX_{j_1}\cdots X_{j_m}
f_{j_{d_1}}\dots f_{j_{d_p}},
\end{equation}
where $(J,B)$ runs over all solutions of \Ref{qt2} such that
$b_1>1$. The exponent $a=a(J,B)$ is \Ref{defa}, and
$\{d_1<\cdots<d_p\}$ is the complement of $B$.
\end{lemma}
\noindent
Our problem is therefore reduced to the combinatorial problem
of finding all solutions $(J,B)$ of \Ref{qt2} with $b_1>1$
and computing their contribution to the trace. The problem can be
reformulated as follows: consider a circle with
$p$ distinct marked points numbered counterclockwise
from 1 to $p$ starting from a base point $*$ distinct
from the marked points. The marked points are assigned
labels: the $j$th point is assigned the label $l_j$.
 Solutions $(J,B)$ are in one-to-one
correspondence with games that a player can play erasing
marked point on this
circle according to the following rules. The player walks
around the circle in {\em clockwise} direction starting from
$*$ a finite number of times to return to $*$. Each move
consists of proceeding to the next (not yet erased) marked
point. When
the player  meets  a
marked point, he or she has the option
of erasing it. If he or she does not erase it, the score is
$X_l$ where $l$ is the label of the point, and we say that
the player ``visited'' the point. The score
for erasing the point is $X_lq^{(\alpha_l,\beta)}$, where
$\beta=\sum_jm_j\alpha_{l_j}$ and $m_j$ is the number of times
the $j$th point has been visited previously. The
game continues until all marked points have been erased. The score
of the game is the product of scores collected during the game.
For instance, if $L=(1,1,2)$, one game consists of the sequence
VEVEE of visits (V) and erasures (E). Its score is
\be X_2X_1q^{(\alpha_1,\alpha_2)}
X_1X_2q^{(\alpha_2,\alpha_1+\alpha_2)}X_1q^{(\alpha_1,\alpha_1+\alpha_2)}
=
X_1^3X_2^2q^{(\alpha_1,\alpha_2)+|\alpha_1+\alpha_2|^2}.
\ee
The relation with the previous description is that
$j_m,\dots,j_2,j_1$ is the list of labels of
marked points met (visited or erased) at each move
of the game, and $B$ indicates at which moves points are visited.
The contribution of a solution $(J,B)$ to \Ref{qt3}
is the overall factor
\begin{equation}\label{overall}
q^{-(\sum_j\alpha_{l_j},\xi-\rho-\sum_j\alpha_{l_j})}e^{2\pi i\xi(\lambda)}
\end{equation}
times the score of the corresponding game times $f_{i_p}\cdots f_{i_1}$
where $i_1,\dots,i_p$ are the labels of the erased points, in order
of erasure.

The next step is a reduction to the classification of
``minimal'' games. An {\em empty round} in a game is a sequence of
of subsequent visits of all marked points exactly once
 without any erasures. A game without empty rounds is called
{\em minimal}. Out of a game we can construct a new
game by inserting an empty round just before an erasure. The
score of the new game is $\prod_{s\in E} X_{l_s}q^b$ times the score of
of the old game, where the product is taken over the set $E$ of
 marked points yet to be erased, and $b=|\sum_{s\in E}\alpha_{l_s}|^2$.
Any game can be obtained from a unique minimal game by
doing this construction sufficiently many times. The games obtained
this way have all the same sequence of erased points, and give
thus proportional contributions to the trace.

Minimal games are in one-to-one correspondence with
permutation $\sigma\in S_p$: we erase first $\sigma(p)$,
then $\sigma(p-1)$, and so on, always at the earliest opportunity.
The example above gives the minimal game corresponding to the
permutation (231). If $\sigma$ is the identity, the score of
the corresponding minimal game is $X_{l_1}\cdots X_{l_p}$.
For general $\sigma$, we must calculate the numbers $b_{jk}$
of times the point $\sigma(k)$ has been visited before
$\sigma(j)$ is erased. The score of the minimal game corresponding to
$\sigma$ is then
\begin{equation}\label{qt9}
 q^c\prod_{j=1}^pX_{l_{\sigma(j)}}^{b_{jj}+1},\qquad
c=\sum_{j,k=1}^pb_{jk}(\alpha_{l_{\sigma(j)}},
\alpha_{l_{\sigma(k)}}).
\end{equation}
The sum of the scores of all games obtained from this game by inserting
empty rounds is then
\begin{equation}\label{score}
{q^c\prod_{j=1}^pX_{l_{\sigma(j)}}^{b_{jj}+1}\over
\prod_{j=1}^p\bigl(1-X_{l_{\sigma(1)}}\cdots X_{l_{\sigma(j)}}
q^{|\alpha_{l_{\sigma(1)}}+\cdots+\alpha_{l_{\sigma(j)}}|^2}\bigr)}.
\end{equation}
We proceed to compute the numbers $b_{jk}$. Roughly speaking,
$b_{jk}$ is the number of times one goes around the circle before
erasing either $\sigma(j)$ or $\sigma(k)$. More precisely,
let
\be
S^{\times}=\{m\in\{1,\dots,p-1\}|\sigma(m)>\sigma(m+1)\}.
\ee
If $k<j$ and $\sigma(k)>\sigma(j)$, $\sigma(j)$ is erased first,
and $b_{jk}$ is the number of times the player passes through the
base point $*$, including at the beginning of the game, before
erasing $\sigma(j)$. This number is
$b_{jk}=|\{m\in S^\times| m\geq j\}|+1$. If $k<j$, and
$\sigma(k)<\sigma(j)$, $b_{jk}$ is the number of
passages through $*$ before erasing $\sigma(j)$, minus 1: $b_{jk}=
|\{m\in S^\times| m\geq j\}|$. If $k\geq j$, we get in
all cases the number of passages through $*$ before erasing
$\sigma(k)$, minus 1:
$b_{jk}=|\{m\in S^\times| m\geq k\}|$.
In particular, $b_{jj}=|\{m\in\{j,\dots,p-1\}|\sigma(m)>\sigma(m+1)\}$.
The exponent $c$ of $q$ in \Ref{qt9} is then
\be
c=\sum_{k<j,\sigma(k)>\sigma(j)}
(\alpha_{l_{\sigma(j)}},\alpha_{l_{\sigma(k)}})
+\sum_{m\in S^\times}(\sum_{j=1}^m\alpha_{l_{\sigma(j)}},
\sum_{j=1}^m\alpha_{l_{\sigma(j)}}).
\ee
The formula in the theorem is then obtained by combining this
formula with \Ref{score}, \Ref{overall}.

\section*{Acknowledgement}
We are grateful to Pavel Etingof for various useful explanations
and suggestions.


\begin{thebibliography}{99}
\bibitem{CP} V. Chari and A. Pressley,
{\it A guide to quantum groups},
Cambridge University Press, 1994
\bibitem{CV} O. Chalykh and A. Veselov,
{\it Commutative rings of partial differential operators and Lie alrebras.}
Commun. Math. Phys. 126 (1990), 597--611
\bibitem{EK1} P. Etingof and A. Kirillov, Jr.,
{\it A unified representation-theoretic approach to special functions}
Funct.\ Anal.\ Appl. 28:1 (1994), 91--94
\bibitem{EK2} P. Etingof and A. Kirillov, Jr.,
{\it Macdonald's polynomials and representations of quantum groups}
Math. Res. Lett. 1 (1994), 279--296
\bibitem{EK3}  P. Etingof and A. Kirillov, Jr.,
{\it Representation-theoretic proof of inner product and symmetry
identities for Macdonald's polynomials},
 hep-th/9410169, to appear in Comp. Math.
\bibitem{ES} P. Etingof and K. Styrkas,
{\it Algebraic integrability of Schr\"odinger operators
ans representations of Lie algebras},
hep-th 9403135 (1994), to appear in Comp.\ Math.
\bibitem{FV}
G. Felder and A. Varchenko,
{\it Integral representation of solutions of the elliptic
Knizhnik--Zamolodchikov--Bernard equations}, Int. Math. Res. Notices
No.\ 5 (1995), 221--133, and paper in
preparation.
\bibitem{HOI} G. Heckman and E. Opdam,
{\it Root systems and hypergeometric functions I},
 Comp. Math. 64 (1987), 329--352
\bibitem{He} S. Helgason,
{\it Groups and geometric analysis},
Associated Press, 1984
\bibitem{Hu} J. Humphreys,
{\it Introduction to Lie algebras and representation theory},
Springer, 1972.
\bibitem{KR} P. Kulish, N. Reshetikhin and E. Sklyanin,
{\it Yang--Baxter equation and representation theory I},
Lett.\ Math.\ Phys.\ 5 (1981), 393--403
\bibitem{Mac} I. Macdonald,
{\it A new class of symmetric functions},
Publ.\ IRMA Strasbourg, 372/S-20, S\'eminaire Lotharingien (1988),
131--171
\bibitem{OT} P. Orlik and H. Terao,
{\it The number of critical points of a product of powers of linear
factors},
Inv.\ Math. 120 (1995), 1-14
\bibitem{RV} N. Reshetikhin and A. Varchenko,
{\it Quasiclassical asymptotics of solutions to the KZ equations},
in ``Geometry, topology, and physics, for Raoul Bott'', S.-T.\ Yau, ed.,
International Press 1995,  293--322
\bibitem{SV} V. Schechtman and A. Varchenko,
{\it Arrangements of hyperplanes and Lie algebra homology},
 Inv. Math. 106 (1991), 139--194
\bibitem{S} B. Sutherland
{\it Exact results for a quantum many-body problem in one dimension},
Phys. Rev. A4  (1971), 2019--2021; Phys. Rev. A5 (1972), 1372--1376
\bibitem{V} A. Varchenko,
{\it Multidimensional hypergeometric functions and representation theory
of quantum groups},
Advanced series in mathematical physics, vol.\ 21,
World Scientific, 1995
\bibitem{V1} A. Varchenko,
{\it Critical points of the product of powers of linear functions and
families of bases of singular vectors},
Comp.\ Math.\ 97 (1995), 385--401
\bibitem{WW}  E. T. Whittaker and G. N. Watson,
{\it Modern analysis}, Cambridge University Press, 1927
\end{thebibliography}
\end{document}